\documentclass[%
 reprint,
superscriptaddress,
showpacs,
 amsmath,amssymb,
aps,
pra,
]{revtex4-1}

\usepackage{graphicx}
\usepackage{dcolumn}
\usepackage{bm}

\usepackage{braket}
\usepackage{color}
\usepackage[caption=false]{subfig}

\usepackage{mathrsfs}

\begin{document}

\preprint{APS/123-QED}

\title{Superradiant and Dark States in Non-Hermitian Plasmonic Antennas and Waveguides}

\author{Amin Tayebi}
\email{tayebiam@msu.edu}
\affiliation{Department of Electrical and Computer Engineering, College of Engineering, Michigan State University, East Lansing, Michigan 48824, USA}
\affiliation{Department of Physics and Astronomy, Michigan State University, East Lansing, Michigan 48824, USA}
\author{Scott Rice}%
\thanks{scott.a.rice@gmail.com }
\affiliation{Department of Electrical and Computer Engineering, College of Engineering, Michigan State University, East Lansing, Michigan 48824, USA}
\affiliation{Department of Physics and Astronomy, Michigan State University, East Lansing, Michigan 48824, USA}

\date{\today}

\begin{abstract}
One-dimensional structures of non-Hermitian plasmonic metallic nanospheres are studied in this paper. For a single sphere, solving Maxwell's equations results in quasi-stationary eigenmodes with complex quantized frequencies. Coupled mode theory is employed in order to study more complex structures. The similarity between the coupled mode equations and the effective non-Hermitian Hamiltonians governing open quantum systems allows us to translate a series of collective phenomenon emerging in condensed matter and nuclear physics to the system of plasmonic spheres. A nontrivial physics emerges as a result of strong non-radiative near field coupling between adjacent spheres. For a system of two identical spheres, this occurs when the width of the plasmonic resonance of the uncoupled spheres is twice the imaginary component of the coupling constant. The two spheres then become coupled through a single continuum channel and the effect of coherent interaction between the spheres becomes noticeable. The eigenmodes of the system fall into two distinct categories: superradiant states with enhanced radiation and dark states with no radiation. The transmission through one-dimensional chains with an arbitrary number of spheres is also considered within the effective Hamiltonian framework which allows us to calculate observables such as the scattering and transmission amplitudes. This nano-scale waveguide can undergo an additional superradiance phase transition through its coupling to the external world. It is shown that perfect transmission takes place when the superradiance condition is satisfied.
\end{abstract}

\maketitle


\section{Introduction} \label{introduction}

Manipulation of light in nanometer scales via surface plasmonic resonances of metallic structures has attracted a great deal of attention over the past two decades \cite{intro1}. Optical antennas capable of localizing light in sub-wavelength regions have resulted in a new generation of photonic devices with applications ranging from imaging \cite{intro2, intro2p5, intro3} to biosensing \cite{intro4,intro5} and emission enhancement of photon sources \cite{intro6,intro7,intro8}. In addition, plasmonics ought to play an important role in the efficient reception and transport of optical energy in light harvesting devices \cite{intro15}. Therefore, various waveguide structures are being investigated in order to control and further improve the propagation of light in micrometer length scales \cite{intro13, intro14,Plasmonics.9.925,PhysRevB.82.035434}.

More recently, it has been shown that surface plasmons can exhibit quantum interference \cite{intro9,intro10}. This has sparked a great interest in studying the quantum properties of surface plasmons and exploiting plasmonic devices as potential building blocks of quantum computers and quantum circuits \cite{intro11,intro12}.  

It was previously suggested that plasmonic structures could be mapped to quantum systems governed by non-Hermitian Hamiltonians \cite{NonhemitianPlasmonic1,NonhemitianPlasmonic2}. In \cite{NonhemitianPlasmonic2}, the radiation properties of an array of optical dipole antennas are manipulated by altering the anti-Hermitian coupling strength between the elements of the array. However, due to the complexity of nano-dipole antennas and the lack of closed form expressions for the fields, the mapping to a non-Hermitian Hamiltonian was achieved via numerical simulation and curve fitting. In this paper we consider systems of plasmonic nano-spheres using the effective non-Hermitian Hamiltonian framework. This framework provides a general platform for studying different physical systems; it has been previously utilized in various problems ranging from quantum signal transmission in nano-structures \cite{celardo09,Greenberg_transport} to solid state quantum computing \cite{Tayebi_qc,PhysRevA.78.062116} and nuclear reactions \cite{VZ777,AZ07,PhysRevLett.115.052501,DrZ_New,PhysRevC.86.044602}. The description of plasmonic structures via the effective Hamiltonian is achieved due to the correspondence between the Feshbach formalism and the coupled mode theory of optical resonators. The effective non-Hermitian Hamiltonian approach allows us to translate phenomena already known in condensed matter and nuclear systems to the plasmonic system under study. One such example is the existence of superradiant and subradiant, or dark, states. In addition, the effective Hamiltonian framework can be readily used in order to calculate observables, such as the transmission coefficient through a plasmonic waveguide.


In Section~\ref{secII} we discuss the effective non-Hermitian Hamiltonian formalism used for studying \emph{open} system in condensed matter \cite{Greenberg_transport,Tayebi_qTrans,Amin_Scott_Conf,YakovG3}, nuclear physics  \cite{Volya_NucP,Volya2014_new,Volya_new2,AUERBACH200445} and quantum optics \cite{YakovG1,YakovG2,YakovG4}. We then discuss the coupled mode formalism used for studying optical and plasmonic systems. The connection between the effective Hamiltonian and coupled mode theory is illustrated through simple two-level examples. We then consider a single plasmonic metallic nanosphere. 

Section~\ref{secIII} considers a single plasmonic metallic nanosphere. The wave equation is solved in order to find the natural resonant frequencies of the single sphere, and discuss the intrinsic radiative nature of nanospheres. This provides a basis for the consideration of two spheres via coupled mode theory. 

The signature of superradiance emerges when the interaction between adjacent optical nano antennas occurs through a single continuum channel, resulting in states with enhanced radiation and confined dark modes. This is discussed in Section~\ref{secIV}. The effect of these states on energy transmission through a one-dimensional chain of spheres is considered in Section~\ref{secV}, with applications to optical frequency nanoscale antennas and waveguide-like structures.

Section~\ref{secVI} includes the summary, concluding remarks and future work.

\section{The effective non-hermitian hamiltonian and coupled mode theory} \label{secII}

In this section we briefly discuss the effective non-Hermitian Hamiltonian approach and the coupled mode theory and show how both theories result in matrices of similar structures. 

\subsection{The Effective Hamiltonian}

Consider a quantum system described by the Hamiltonian $H_0$ and its discrete set of eigenvectors $\ket{i}$ that interacts with its surrounding environment. The environment is thermodynamically large and is characterized by infinitely many channels with continuous energy spectrum $\ket{c;E}$. The Hilbert space of this problem can be divided into two subspaces with the help of projection operators $\mathcal{Q}$ and $\mathcal{P}$. The operator $\mathcal{Q}$ only acts on the subspace of the closed system and the operator $\mathcal{P}$ acts on the environment only. Without loss of generality, one can further assume that the projections are orthogonal, therefore $\mathcal{P}+\mathcal{Q}=1$ and $\mathcal{P}\mathcal{Q}=\mathcal{Q}\mathcal{P}=0$. The total Hamiltonian $H$ i.e. the Hamiltonian of the system, the environment and their interactions, is then decomposed into $H=H_{\mathcal{Q}\mathcal{Q}}+H_{\mathcal{Q}\mathcal{P}}+H_{\mathcal{P}\mathcal{Q}}+H_{\mathcal{P}\mathcal{P}}$, where $H_{\mathcal{Q}\mathcal{Q}}=\mathcal{Q}H\mathcal{Q}$, $H_{\mathcal{Q}\mathcal{P}}=\mathcal{Q}H\mathcal{P}$, $H_{\mathcal{P}\mathcal{Q}}=\mathcal{P}H\mathcal{Q}$ and $H_{\mathcal{P}\mathcal{P}}=\mathcal{P}H\mathcal{P}$.  The effective Hamiltonian is achieved by projecting the stationary wave function in the Schr\"{o}dinger equation $H\Psi=E\Psi$ into the subspace of the closed system
\begin{equation} \label{effectiveHamiltonian1}
\mathscr{H}_{\text{eff}} (E) \mathcal{Q} \Psi = E\mathcal{Q} \Psi,
\end{equation}
where 
\begin{equation} \label{effectiveHamiltonian2}
\mathscr{H}_{\text{eff}} (E) =  H_{\mathcal{Q}\mathcal{Q}} + H_{\mathcal{Q}\mathcal{P}}  \frac{1}{E-H_{\mathcal{P}\mathcal{P}} } H_{\mathcal{P}\mathcal{Q}} .
\end{equation}
The first term in the left handside of (\ref{effectiveHamiltonian2}) is the Hamiltonian of the closed system $\mathcal{Q} H \mathcal{Q}=H_0$. The second term in (\ref{effectiveHamiltonian2}), can be further simplified by calculating the matrix element between two intrinsic states of the closed system $\ket{i}$ and $\ket{j}$
\begin{equation} \label{effectiveHamiltonian3}
\langle i| H_{\mathcal{Q}\mathcal{P}}  \frac{1}{E-H_{\mathcal{P}\mathcal{P}} } H_{\mathcal{P}\mathcal{Q}} \ket{j}= \sum_{c}\int{dE'\frac{A_{i}^{c}(E'){A_{j}^{c}}^{*}(E')}{E-E'}},
\end{equation}
where $A_{i}^{c}(E)$ are transition amplitudes from continuum channel $\ket{c;E}$ to internal state $\ket{i}$; $A_{i}^{c}(E)=\langle i|H_{\mathcal{Q}\mathcal{P}}\ket{c;E}$. Usually continuum channels couple to the system only if the energy is above a certain threshold in which case the channel is open, otherwise the channel is closed and the transition amplitude vanishes. Using the Sokhotski-Plemelj theorem, the integral in (\ref{effectiveHamiltonian3}) can be decomposed into Hermitian and anti-Hermitian parts
\begin{equation}\label{effectiveHamiltonian4}
   \vspace*{-16mm} \sum_{c} \int{dE'}\frac{A_{i}^{c}(E'){A_{j}^{c}}^{*}(E')}{E-E'} = \\
   \end{equation}
   \vspace{11mm}
   \begin{equation}
   \sum_{c}\mathcal{P}.\mathcal{V}.\int{dE'\frac{A_{i}^{c}(E'){A_{j}^{c}}^{*}(E')}{E-E'}} 
  -  i \pi \sum_{c_{open}} A_{i}^{c}(E){A_{j}^{c}}^{*}(E), \nonumber
\end{equation} 
where P.V. denotes the Cauchy principal value. Accordingly, two operators are defined; $\Delta(E)$, corresponding to the Hermitian component in (\ref{effectiveHamiltonian4}) with matrix elements
\begin{equation} \label{realpartHeff}
\Delta_{i,j}(E) = \sum_{c}\mathcal{P}.\mathcal{V}.\int{dE'\frac{A_{i}^{c}(E'){A_{j}^{c}}^{*}(E')}{E-E'}},
\end{equation}
and $W(E)$ corresponding to the anti-Hermitian component in (\ref{effectiveHamiltonian4}), with matrix elements
\begin{equation} \label{imagpartofHeff}
W_{ij}(E)= 2\pi \sum_{c_{open}} A_{i}^{c}(E){A_{j}^{c}}^{*}(E).
\end{equation}
Thus, in operator form, the \emph{energy-dependent} effective Hamiltonian is
\begin{equation}  \label{effectiveHamiltonian_complete}
\mathscr{H}_{\text{eff}} (E)= H_0+\Delta(E)-\frac{i}{2}\,W(E).
\end{equation}
The two terms, $\Delta(E)$ and $W(E)$, also known as the \emph{self energy}, completely take the effect of the interaction with the environment into account. The Hermitian part, $\Delta(E)$, renormalizes the energies of the closed system. Notice that the summation in (\ref{realpartHeff}) runs over all continuum channels, open and close. This is because even when the running energy, $E$, is below the threshold such that $A_{i}^{c}(E)$ vanishes, the integral is still non-vanishing and thus takes the \emph{virtual} coupling to the continuum into account. On the other hand, $W(E)$ which is responsible for the decay width of the energy states of the effective Hamiltonian (\ref{effectiveHamiltonian_complete}), arises only due to real interaction processes with the environment.

In many situations, including the case we are concerned with in this paper, the energy window of interest is relatively narrow and the transition amplitudes $A_{i}^{c}(E)$ are smooth functions of energies. These amplitudes can therefore be considered as energy-independent quantities. Consequently, the integral in the Hermitian component of the self energy (\ref{realpartHeff}) vanishes and the effective Hamiltonian reduces to
\begin{equation} \label{ReducedHeff}
\mathscr{H}_{\text{eff}}=H_{0}-\frac{i}{2}\,W.
\end{equation}

It is interesting to look at the statistics of the \emph{quasi-stationary} eigenenergies of the effective Hamiltonian
\begin{equation} \label{Heff_energies}
E_{n}=\mathrm{E}_{n}-\frac{i}{2}\Gamma_n,
\end{equation}
where $\mathrm{E}_{n}$ is the real component of the energy and  $\Gamma_n$ is the width of the state and related to the lifetime by $\tau_n=\hbar/\Gamma_n$. The positiveness of $\Gamma_n$ is guaranteed by the Cholesky factorized form of $W$ in (\ref{imagpartofHeff}) which makes $W$ a positive definite matrix. We define the quantity $\xi$ that parameterizes the strength of interaction with the external world, thus
\begin{equation} \label{ReducedHeff_xi}
\mathscr{H}_{\text{eff}}=H_{0}-\frac{i}{2}\xi\,W.
\end{equation}
Because the framework is exact and no approximation was used, $\xi$ can take arbitrarily small values representing weak interactions or extremely large values representing strong interactions with the external world. For weak interactions, when $\xi$ is small, the anti-Hermitian component, $W$ is a perturbation to $H_0$. The complex eigenenergies are then narrow resonances  with almost uniform width distribution \cite{SOKOLOV1,Volya2016}. In the opposite limit of strong interactions, the anti-Hermitian component becomes the dominant term and $H_0$ is the perturbation. It is clear from (\ref{imagpartofHeff}) that the rank of $W$ is equal to the number of open channels which is normally much smaller than the number of intrinsic states of the closed system. Therefore a few states become dominant resonances that consume the entire width and the remaining states become long-lived states that decouple from the environment. Due to the resemblance of this phenomenon to the Dicke superradiance in quantum optics, we term the broad short-lived resonances as superradiant states and the narrow short-live resonances as subradiant states. It was shown \cite{Tayebi_qc,Amin_Thesis} that if the system is connected to the environment through its physical boundaries then the superradiant states are localized to the boundaries of the system, and the subradiant states are pushed away from the boundaries and trapped within the interior of the system. The superradiance \emph{phase transition} is discussed more rigorously in \cite{SOKOLOV1,Auerbach}. It was shown that  the important parameter to consider is the ratio of $\langle \Gamma \rangle$, the average widths of the energies in (\ref{Heff_energies}), to $D$, the mean level spacing of the close system. The transition occurs when $\langle \Gamma \rangle / D \approx 1$, this is when the resonances are maximally overlapped and the superradiant states emerge.

The effective non-Hermitian Hamiltonian framework also provides us with useful expressions corresponding to observables such as the scattering and transmission amplitudes \cite{SOKOLOV1,Auerbach,SOKOLOV2,Amin_Thesis}. Here we are particularly interested in transmission through a one dimensional chain of metallic nano-spheres. The transmission amplitude from a continuum channel $a$ to channel $b$ through the system is given by
\begin{equation} \label{Transmission_amp}
Z^{ab}(E)=\sum_{i,j} A_{i}^{a} \bigg( \frac{1}{E-\mathscr{H}_{\text{eff}}} \bigg)_{i,j} {A_{j}^{b}}^*.
\end{equation} 
The transmission probability, $T^{ab}(E)$ is therefore
\begin{equation} \label{ProcessAmplitude}
T^{ab}(E)=|Z^{ab}(E)|^2.
\end{equation}
The amplitude in (\ref{Transmission_amp}) has a simple interpretation: the particle enters into state $\ket{j}$ through channel $b$. Next the propagator takes the particle from state $\ket{j}$ to $\ket{i}$ considering all possible \emph{paths}. Finally the particle escapes the system from site $\ket{i}$ through continuum channel $a$.

\subsection{Coupled Mode Theory}

In this paper, we employ coupled mode theory in order to study systems of interacting plasmonic spheres. This method is reminiscent of the time-dependent perturbation theory in quantum mechanics; it has been used for the investigation of coupled resonators in optical systems \cite{CMT_optic1,CMT_optic2,CMT_optic3}, wireless energy transfer loop antennas \cite{CMT_antennas1}, and plasmonic structures including antennas \cite{NonhemitianPlasmonic2,CMT_antennas1} and waveguides \cite{CMT_waveguide1,CMT_waveguide2,CMT_waveguide3}. The coupled mode approach significantly simplifies the complexity of the problem: instead of solving the wave equation one needs to solve a system of linear algebraic equations. In addition, it provides a clear and intuitive picture of how interactions between the constituents of the system can dramatically change the dynamics.

The formulation provided in this paper, similar to \cite{cmt_qnm_new}, is rather general. No specific boundary conditions are assumed and hence it is applicable to the system of coupled plasmonic nanoantennas discussed in the future sections. 

Consider two non-magnetic dielectric resonators with relative dielectric constants $\epsilon_1(\vec{r})$ and $\epsilon_2(\vec{r})$. The resonators occupy a volume in space, $V_1$ and $V_2$, respectively. In addition, consider that the relative dielectric constants $\epsilon_1(\vec{r})$ and $\epsilon_2(\vec{r})$ are equal to unity for points outside of the resonators. In a time harmonic scenario each resonator, when isolated, satisfies the wave equation
\begin{equation} \label{wave_eqn_cmt}
\vec{\nabla} \times \vec{\nabla} \times \vec{\mathcal{E}}^{\alpha}_n(\vec{r})-\bigg(\frac{ \omega_{\alpha , n}}{c}\bigg)^2   \epsilon_{\alpha} (\vec{r}) \vec{\mathcal{E}}^{\alpha}_n(\vec{r})=0, 
\end{equation}
where $\alpha=1, 2$ denotes the resonator number and $c$ is the speed of light in the background medium which is assumed to be the free space for simplicity. Due to the sharp discontinuity between the resonator and the background at the resonator boundaries, the modes are quantized and characterized by the integer number $n=1, 2, 3, ...$ and their eigenfrequency $\omega_{\alpha,n}$. The modes of isolated resonators are normalized according to 
\begin{equation} \label{normalization_isolated_mode_gen}
\int_{V}  \epsilon_{\alpha}(\vec{r}) {\vec{\mathcal{E}}^{\alpha^*}_m}(\vec{r}). \vec{\mathcal{E}}^{\alpha}_n(\vec{r}) d^3r = \delta_{mn},
\end{equation}
where $V$ is the total volume in which fields are present and $m$ and $n$ are the mode indices and $\delta_{mn}$ is the Kronecker delta. The normalization expression is of crucial importance and its form is dictated by the boundary conditions of the problem. For instance in \cite{cmt_qnm_new} the normalization is similar to (\ref{normalization_isolated_mode_gen}), however with no complex conjugation. As we will see, the normalization expression has to be modified when discussing spherical plasmonic particles in the following sections, in order for the normalization expression to remain finite when integrated over all space. For now, we assume that the normalization rule is given by the general dot product definition provided in (\ref{normalization_isolated_mode_gen}) with the integration volume being all space, as this is usually the case in electromagnetic textbooks. We will revisit the normalization definition later in this paper. 

Next we assume that, for a system of two coupled resonators, the total electric field, $\vec{\mathcal{E}}(\vec{r})$, can be written as a superposition of a finite number of individual modes of the two resonators:
\begin{equation} \label{ansatz_cmt}
\vec{\mathcal{E}}(\vec{r})=\sum_{n=1}^{N} \Big[ a_1(n) \vec{\mathcal{E}}^{1}_n(\vec{r}) + a_2(n) \vec{\mathcal{E}}^{2}_n(\vec{r})\Big],
\end{equation}
where $N$ is the total number of modes. The total electric field satisfies the wave equation 
\begin{equation} \label{wave_eqn_cmt_total_field}
\vec{\nabla} \times \vec{\nabla} \times \vec{\mathcal{E}}_n(\vec{r})-\Big(\frac{\omega_n}{c}\Big)^2 \epsilon(\vec{r}) \ \vec{\mathcal{E}}_n(\vec{r})=0, 
\end{equation}
where $\omega_n$ are the eigenfrequencies of the coupled system and $\epsilon(\vec{r})$ is the dielectric constant at a given point, $\vec{r}$, when both resonators are simultaneously present. The function $\epsilon(\vec{r})$ is equal to $\epsilon_1(\vec{r})$ and $\epsilon_2(\vec{r})$ for points inside the first and second resonator, respectively, and is equal to unity otherwise. Plugging the ansatz (\ref{ansatz_cmt}) into the wave equation (\ref{wave_eqn_cmt_total_field}) and using its linearity and (\ref{wave_eqn_cmt}) we arrive at
\begin{align} \label{series_eqn}
 \sum_{n=1}^N \Big[ &  a_1(n) \big(\omega_{1,n}\big)^2 \epsilon_{1} (\vec{r}) \vec{\mathcal{E}}^{1}_n(\vec{r}) + a_2(n) \big(\omega_{2,n}\big)^2 \epsilon_{2} (\vec{r}) \vec{\mathcal{E}}^{2}_n(\vec{r}) \Big] \nonumber \\
& =\omega_n^2 \epsilon (\vec{r}) \sum_{n=1}^N \Big[  a_1(n) \vec{\mathcal{E}}^{1}_n(\vec{r}) + a_2(n) \vec{\mathcal{E}}^{2}_n(\vec{r}) \Big].
\end{align}
Using the normalization rule (\ref{normalization_isolated_mode_gen}) to project (\ref{series_eqn}) onto $\vec{\mathcal{E}}^{1}_m(\vec{r})$ and $ \vec{\mathcal{E}}^{2}_m(\vec{r})$ for all values of $m$: $m=1,2,..,N$, we obtain a system of $2N$ linear equations. In the matrix form
\begin{equation} \label{cmt_equation_complete_form}
\begin{pmatrix}
    \bm{T^{11}} & \bm{T^{12}}  \\
    \bm{T^{21}} & \bm{T^{22}}  
  \end{pmatrix}
 \begin{pmatrix}
    \bm{\Omega_{1}}^2 & \bm{0}  \\
    \bm{0} & \bm{\Omega_{2}}^2  
  \end{pmatrix}
  \begin{bmatrix}
    \vec{A}_1 \\ 
    \vec{A}_2
  \end{bmatrix}= 
  \omega^2 
  \begin{pmatrix}
    \bm{L^{11}} & \bm{L^{12}}  \\
    \bm{L^{21}} & \bm{L^{22}}  
  \end{pmatrix}
  \begin{bmatrix}
    \vec{A}_1 \\ 
    \vec{A}_2
    \end{bmatrix}, 
\end{equation}
where $\vec{A}_1$ and $\vec{A}_2$ are $N \times 1$ vectors of the coefficients $a_1(n)$ and $a_2(n)$ in the ansatz (\ref{ansatz_cmt}), respectively. $\bm{\Omega_{1}}$ and $\bm{\Omega_{2}}$ are $N \times N$ diagonal matrices containing the eigenfrequencies of the isolated resonators with matrix elements
\begin{equation}
\big(\Omega_{\alpha}\big)_{mn}= \omega_{\alpha,n} \ \delta_{mn}
\end{equation}
where as previously $\alpha=1, 2$. The matrix elements of the four square $N \times N$ matrices $\bm{T^{\alpha \beta}}$, where $\alpha, \beta=1, 2$, are given by
\begin{equation}
T^{\alpha\beta}_{mn}= \int_{V}  \epsilon_{\beta}(\vec{r}) {\vec{\mathcal{E}}^{\alpha^*}_m}(\vec{r}). \vec{\mathcal{E}}^{\beta}_n(\vec{r}) d^3r. 
\end{equation}
According to (\ref{normalization_isolated_mode_gen}), $\bm{T^{11}}$ and $\bm{T^{22}}$ are equal to the identity matrix $\bm{1}$. Finally the elements of the matrices $\bm{L^{\alpha, \beta}}$ are given by
\begin{equation} \label{L_matrix_cmt}
L^{\alpha\beta}_{mn}=\int_{V}  \epsilon(\vec{r}) {\vec{\mathcal{E}}^{\alpha^*}_m}(\vec{r}). \vec{\mathcal{E}}^{\beta}_n(\vec{r}) d^3r. 
\end{equation}
Because the dielectric function $\epsilon(\vec{r})$ is the sum of the two dielectric function, the matrix elements $L^{12}_{m,n}$ and $T^{12}_{m,n}$ are related via
\begin{equation}  \label{T_matrix_cmt}
L^{12}_{mn}=T^{12}_{mn}+\int_{V_1} \big(\epsilon_1(\vec{r})-1 \big) \vec{\mathcal{E}}^{1^{*}}_m(\vec{r}).\vec{\mathcal{E}}^{2}_n(\vec{r}) d^3r,
\end{equation}
where the integration is carried out over the volume of the first resonator, $V_1$ only. Accordingly, we define the matrix $\bm{K^{12}}$ with matrix elements
\begin{equation} \label{coupling_coeff_K1}
K^{12}_{mn}=\int_{V_1} \big(\epsilon_1(\vec{r})-1 \big) \vec{\mathcal{E}}^{1^{*}}_m(\vec{r}).\vec{\mathcal{E}}^{2}_n(\vec{r}) d^3r.
\end{equation}
Therefore
\begin{equation} \label{kappa_matrix1}
\bm{L^{12}}=\bm{T^{12}}+\bm{K^{12}}. 
\end{equation}
Similarly $L^{21}_{mn}$ is related to $T^{21}_{mn}$ as
\begin{equation}
L^{21}_{mn}=T^{21}_{mn}+\int_{V_2} \big(\epsilon_2(\vec{r})-1 \big) \vec{\mathcal{E}}^{2^{*}}_m(\vec{r}).\vec{\mathcal{E}}^{1}_n(\vec{r}) d^3r.
\end{equation}
Correspondingly $\bm{K^{21}}$ is defined with matrix elements
\begin{equation} \label{coupling_coeff_K2}
K^{21}_{mn}=\int_{V_2} \big(\epsilon_2(\vec{r})-1 \big) \vec{\mathcal{E}}^{2^{*}}_m(\vec{r}).\vec{\mathcal{E}}^{1}_n(\vec{r}) d^3r.
\end{equation}
Hence
\begin{equation} \label{kappa_matrix2}
\bm{L^{21}}=\bm{T^{21}}+\bm{K^{21}}. 
\end{equation}

In order to simplify (\ref{cmt_equation_complete_form}), we accept a number of approximations that are commonly used in studying systems of weakly coupled resonators \cite{Elnaggar_app,Elnaggar_CMT1,Elnaggar_CMT2}. We assume that the diagonal matrix elements in (\ref{L_matrix_cmt}) are approximately equal to unity, i.e. $L^{11}_{m,n}=L^{22}_{m,n} \approx 1$ and therefore $\bm{L^{11}}=\bm{L^{22}}\approx \bm{1}$. This is justified due to the strong field confinement within the dielectric regions. Furthermore, we assume that the coupling is weak and therefore the coupling elements in (\ref{T_matrix_cmt}) satisfy the condition $T^{12}_{mn} T^{21}_{m'n'} \ll 1$. Using these approximations along with (\ref{kappa_matrix1}) and (\ref{kappa_matrix2}), the coupled mode equation (\ref{cmt_equation_complete_form}) reduces to
\begin{equation} \label{cmt_equation_reduced_form1}
\begin{pmatrix}
    \bm{1} & -\bm{K^{12}}  \\
    -\bm{K^{21}} & \bm{1}  
  \end{pmatrix}
 \begin{pmatrix}
    \bm{\Omega_{11}^2} & \bm{0}  \\
    \bm{0} & \bm{\Omega_{22}^2}  
  \end{pmatrix}
  \begin{bmatrix}
    \vec{A}_1 \\ 
    \vec{A}_2
  \end{bmatrix}= 
  \omega^2 
  \begin{bmatrix}
    \vec{A}_1 \\ 
    \vec{A}_2
    \end{bmatrix}. 
\end{equation}
It is also helpful to linearize the system of equations (\ref{cmt_equation_reduced_form1}). This can be done by noting that the eigenmodes of the isolated resonators are not far apart and are clustered around their mean value \cite{Haus_CMT}, i.e. $\omega \approx \omega_{\alpha,n}$. Under this approximation $\omega$ and $\omega_{\alpha,n}$ satisfy the following 
\begin{equation}
\omega^2-\big(\omega_{\alpha,n} \big)^2 \approx 2\omega_{\alpha,n}(\omega-\omega_{\alpha,n} ).
\end{equation}
This brings us to the final form of the coupled mode equations 
\begin{equation} \label{cmt_equation_reduced_form2}
\begin{pmatrix}
    \bm{1} & -\frac{1}{2}\bm{K^{12}} \\
    -\frac{1}{2}\bm{K^{21}} & \bm{1}  
  \end{pmatrix}
 \begin{pmatrix}
    \bm{\Omega_{11}} & \bm{0}  \\
    \bm{0} & \bm{\Omega_{22}}  
  \end{pmatrix}
  \begin{bmatrix}
    \vec{A}_1 \\ 
    \vec{A}_2
  \end{bmatrix}= 
  \omega 
  \begin{bmatrix}
    \vec{A}_1 \\ 
    \vec{A}_2
    \end{bmatrix}.
\end{equation}

In the simplest situation when the two resonators are identical and only one mode of an isolated resonator is considered, the electric field of the coupled system can be expressed as $\vec{\mathcal{E}}(\vec{r})= a_1 \vec{\mathcal{E}}^1(\vec{r}) + a_2 \vec{\mathcal{E}}^1(\vec{r})$. According to (\ref{cmt_equation_reduced_form2}) the coupled mode equations are then given by 
\begin{align} \label{CMT2res_timeDomain}
\omega_0 a_1 +  \kappa \ a_2=\omega a_1, \nonumber \\
\omega_0 a_2 + \kappa^*a_1=\omega a_2, 
\end{align} 
where $\omega_0$ is the eigenfrequency of the isolated resonators. Using (\ref{coupling_coeff_K1}) and (\ref{cmt_equation_reduced_form2}), $\kappa$ is given by
\begin{equation} \label{coupling_coefficient}
\kappa=-\frac{1}{2}\omega_0 \int_{\text{V}_1}  \big(\epsilon_1(\vec{r})-1 \big) \vec{\mathcal{E}}^{1^*}(\vec{r}). \vec{\mathcal{E}}^2(\vec{r}) d^3r.
\end{equation} 
The coupling coefficient $\kappa$ has a simple interpretation: it is the interaction energy between the field generated by the second resonator and the dipole moment of the first resonator averaged over one period. The complex conjugation of the coupling coefficient in (\ref{CMT2res_timeDomain}) is dictated by the energy conservation, assuming there is no loss or gain in the system \cite{Haus_CMT}. The eigenfrequencies of the coupled system, which are guaranteed to be real due to the Hermitian form of the equations in (\ref{CMT2res_timeDomain}), are
\begin{equation}
\omega_{\pm}=\omega_0 \pm |\kappa|,
\end{equation}
where the frequencies of the coupled system, $\omega_{+}$ and $\omega_{-}$, correspond to the symmetric eigenstate with $a_1=a_2=1/\sqrt{2}$ and the anti-symmetric eigenstate with $a_1=-a_2=1/\sqrt{2}$, respectively.

One can readily see the similarity between the coupled mode theory and the quantum theory as both are a theory of waves. Equation (\ref{CMT2res_timeDomain}) is the Schr\"{o}dinger equation for a two-level system (a qubit). In quantum mechanical language, $\omega_0$ is the energy of the \emph{unperturbed} states and the off-diagonal matrix element $\kappa$ represents the interaction strength between the two states which is responsible for the level repulsion and avoided crossing of the final \emph{mixed} states. 

An interesting dynamic of the two-level system is the so-called Rabi oscillation. Let the system at time $t=0$ be prepared in the unperturbed state with energy $\omega_1$. Then the probability $P(t)$ to find the system in the same state at time $t$ is \cite{zelevinskybook}
\begin{equation}
P(t)=1-\text{sin}^2\Big(\frac{\omega_R t}{2} \Big),
\end{equation} 
where $\omega_R=\omega_{+}-\omega_{-}=2|\kappa|$ is the Rabi frequency of the excitation oscillating back and forth between the two levels. Because the unperturbed energies of the two states are equal, the probability goes through the minimum, $P=0$, which indicates that the excitation can be completely transferred, leaving no residue in the initial state. The Rabi oscillation was predicted in systems of optical waveguides \cite{Rabi_waveguide} and in coupled ring resonators \cite{Rabi_ringRes}.

We now focus on a more realistic case: two coupled identical dielectric resonators where in general, due to damping and leakage of the resonators, the energy is no longer conserved. Therefore the governing equations need not be Hermitian. In the case of open systems one has to modify the normalization expression (\ref{normalization_isolated_mode_gen}) which leads to an altered expression for $\kappa$. The coupling coefficient in this case is similar to (\ref{coupling_coefficient}) but with no complex conjugation (see \cite{cmt_qnm_new} for detail). The coupled equations (\ref{CMT2res_timeDomain}) are modified to a more general form 
\begin{align} \label{CMT2res_freqDomian}
\omega_0 a_1 + \kappa a_2 &=\omega a_1 \nonumber \\
\omega_0 a_2 + \kappa a_1 &=\omega a_2, 
\end{align} 
where $\omega_0$ can now be complex: $\omega_0=\eta_0-i\gamma_0/2$ representing loss and radiation. Similarly, $\kappa$ is in general complex as well: $\kappa=\kappa'-i\kappa''$ where $\kappa'$ and $\kappa''$ are real numbers. The left hand side of the coupled equations (\ref{CMT2res_freqDomian}) can then be written as the summation of two matrices, a Hermitian matrix, $H'_0$, and an anti-Hermitian matrix $W'$:
\begin{equation} \label{CMT_to_eff_Hamiltonian}
  \begin{pmatrix}
    \omega_0 & \kappa  \\
    \kappa & \omega_0  
  \end{pmatrix}=
  H'_0-\frac{i}{2}W',
\end{equation}
where   
 \begin{equation}  \label{CMT_Matrix_hermitian}
  H'_0=
  \begin{pmatrix}
    \eta_0 & \kappa'  \\
    \kappa' & \eta_0  
  \end{pmatrix},
\end{equation}
and 
\begin{equation}  \label{CMT_Matrix_nonhermitian}
  W'=
  \begin{pmatrix}
    \gamma_0 & 2\kappa''  \\
    2\kappa'' & \gamma_0 
  \end{pmatrix}.
\end{equation}
The similarity between the coupled mode theory and the effective non-Hermitian Hamiltonian formalism becomes apparent by comparing (\ref{ReducedHeff}) and (\ref{CMT_to_eff_Hamiltonian}). The problem of two coupled dielectric resonators is mapped to a two level quantum system where in general each level is coupled to an independent continuum channel. This is because in general the rank of $W'$ is 2, therefore, according to (\ref{imagpartofHeff}), one requires two independent open channels to construct the anti-Hermitian matrix $W'$. In the particular case when $\gamma_0=\pm 2\kappa''$, the rank of $W'$ is equal to unity, therefore the superradiance condition is fulfilled and only one open channel is required to construct the matrix $W'$. In this case, the effective Hamiltonian has two distinct eigenvalues, a purely real eigenvalue or the subradiant state , reminiscent of dark modes in open quantum systems, with eigenfrequency $\eta_0-\kappa'$, and a complex eigenvalue with enhanced radiation properties and eigenfrequency $\eta_0+\kappa'-i\gamma_0$, which is the superradiant state.  

\section{A Single Metallic Sphere} \label{secIII}
In this section we consider a single isolated metallic sphere embedded in a homogeneous background dielectric material. The problem is treated classically by solving Maxwell equations. In the absence of external sources and assuming a harmonic time dependence of the form
\begin{equation} \label{Phase_convention}
\vec{\mathbb{E}}(\vec{r},t)=e^{i\omega t} \vec{\mathcal{E}}(\vec{r}), 
\end{equation}
the governing equation is the well known Helmholtz equation
\begin{equation} \label{Helmholtz_eqn}
\nabla^2 \vec{\mathcal{E}}(\vec{r})-k^2\vec{\mathcal{E}}(\vec{r})=0.
\end{equation}
The wave number $k$ is defined for the interior and exterior regions of the plasmonic sphere according to 
\begin{align}
k = \begin{cases}
k_{\text{in}}=\frac{\omega}{c} \sqrt{\epsilon_{\text{in}}} & r\leq a,\\
k_{\text{out}}=\frac{\omega}{c} \sqrt{\epsilon_{\text{out}}}   & r> a,
\end{cases}
\end{align}
where $c$ is the speed of light in vacuum and $a$ is the radius of the sphere which is located at the origin of the coordinate system. The relative dielectric constants of the metallic sphere and the background medium are denoted by $\epsilon_{\text{in}}$ and $\epsilon_{\text{out}}$, respectively. Plasmonic structures are usually made of noble metals, such as gold and silver, with face-centered cubic lattice, or alkali metals, such as sodium and potassium, with body-centered cubic lattice. Due to their symmetric crystal lattice types, they are isotropic to light and their relative permittivity is characterized by a scalar. This dielectric constant is well described by the Drude-Sommerfeld model
\begin{align} \label{Drude_dielectric_func}
\epsilon_{\text{in}}(\omega)=\epsilon_{\infty}-\frac{\omega_p^2}{\omega^2-i\omega\gamma_s},
\end{align}
where $\omega_p$ is the plasma frequency which is defined by the electron effective mass $m^*$, the vacuum permittivity $\epsilon_0$, electron charge $e$, and electron density $n$; $\omega_p^{2}= ne^2/\epsilon_0m^*$. The loss within the dielectric material due to various processes, such as electron-phonon interaction, impurities and scattering, is incorporated into the relaxation rate $\gamma_s$. The negative sign of this term in the denominator is dictated by the phase convention adopted in (\ref{Phase_convention}). The phenomenological parameter $\epsilon_{\infty}$ accounts for the contribution of the bound electrons to the polarization of the dielectric material. For typical metals the plasma frequency, $\omega_p$, ranges from 3 to 15 eV (700-3600 THz) which mainly falls into the ultraviolet spectrum \cite{Drude_model_plasma_1, Drude_model_plasma_2, Drude_model_plasma_3, Drude_model_plasma_4, Drude_model_plasma_5}. The damping rate $\gamma_s$ is much smaller than the plasma frequency, $\gamma_s \ll \omega_p$, being of the order $10^{-2}-10^{-1}$ eV (2.4-24 THz). Finally, the correction term $\epsilon_{\infty}$ typically ranges from 1 to 10 \cite{Silver_drude}. In an ideal electron gas, $\epsilon_{\infty}=1$ and $\gamma_s=0$, therefore the dielectric function (\ref{Drude_dielectric_func}) reduces to $\epsilon_{\text{in}}(\omega)=1-\omega_p^2/\omega^2$. Below the plasma frequency the dielectric function is negative and the field can not penetrate inside the metal. For frequencies larger than the plasma frequency however, the dielectric constant becomes positive and the fields can penetrate the metal i.e. the metal becomes transparent. 

Using the spherical coordinate system, the solutions of the Helmholtz equation (\ref{Helmholtz_eqn}) for the plasmonic sphere can be divided into two categories: transverse magnetic (TM) modes with no radial magnetic field  and transverse electric (TE) modes with no radial electric field component. In this work we consider the TM modes only. The components of the electric field are given by \cite{harringtonbook}: 
\begin{align} \label{field_exprs}
\mathcal{E}_r&=\zeta \ C(r;a) \ell(\ell+1) \frac{f_{\ell}(kr)}{\epsilon(r)kr} Y_\ell^m(\theta,\phi), \nonumber  \\ 
\mathcal{E}_\theta&=\zeta \ C(r;a)\frac{1}{\epsilon(r)kr} \frac{\partial}{\partial(kr)}\Big(krf_\ell(kr)\Big)\frac{\partial}{\partial \theta} Y_\ell^m(\theta,\phi),  \\ 
\mathcal{E}_\phi&=\zeta \ C(r;a)\frac{1}{\epsilon(r)kr} \frac{\partial}{\partial(kr)}\Big(krf_\ell(kr)\Big)\frac{1}{\sin\theta}\frac{\partial}{\partial \phi}Y_\ell^m(\theta,\phi), \nonumber
\end{align}
where $\zeta$ is the normalization constant discussed in detail in the next section. The dielectric constant is equal to $\epsilon_{\text{in}}$ and $\epsilon_{\text{out}}$ for $r\leq a$ and $r>a$, respectively. $Y_\ell^m(\theta,\phi)$ are the spherical harmonics with $\ell=0,1,2,...$ and $m=0,\pm 1,\pm 2,...,\pm \ell$. The case of $\ell=0$ results in the trivial solution. The first non-trivial solution corresponds to the dipole mode, $\ell=1$. The function $f_{\ell}(kr)$ is equal to the spherical Bessel function of the first kind and the spherical Hankel function of the second kind, for $r\leq a$ and $r>a$, respectively.
\begin{align}
f_{\ell}(kr) = \begin{cases}
j_{\ell}(k_{\text{in}}r) & r\leq a,\\
h_{\ell}^{(2)}(k_{\text{out}}r)  & r> a.
\end{cases}
\end{align}
The Bessel function $j_{\ell}(k_{\text{in}}r)$ represent standing waves within the plasmonic sphere while, noting the phase convention (\ref{Phase_convention}), the Hankel function $h_{\ell}^{(2)}(k_{\text{out}}r)$ describes radially outward traveling waves which satisfy the Sommerfeld radiation boundary condition.
The coefficient $C(r;a)$ guarantees that the boundary conditions are satisfied at the boundary of the sphere (see \cite{SingleSph1} for details)
\begin{align} \label{general_constant}
C(r;a) = \begin{cases}
 \big[ \ j_{\ell}(k_{\text{in}}a)\big]^{-1} & r\leq a,\\
 \big[ \ h_{\ell}^{(2)}(k_{\text{out}}a)\big]^{-1}  & r> a.
\end{cases}
\end{align}
Matching the interior and the exterior fields leads to the characteristic equation of the discrete eigenfrequencies of the system:
\begin{equation} \label{charac_eqn_single_sphere}
\epsilon_{\text{in}}\bigg[1+k_{\text{out}}a \ \frac{h_{\ell}^{(2)'}(k_{\text{out}}a)}{h_{\ell}^{(2)}(k_{\text{out}}a)} \bigg]=\epsilon_{\text{out}}\bigg[1+k_{\text{in}}a \ \frac{j'_{\ell}(k_{\text{in}}a)}{j_{\ell}(k_{\text{in}}a)} \bigg].
\end{equation}
Here, the prime denotes differentiation with respect to the argument of the function, i.e. $j'_{\ell}(k_{\text{in}}r)= \partial j_{\ell}(k_{\text{in}}r)/ \partial (k_{\text{in}}r )$. For a given radius, different modes can be labeled by $\ell$: $\omega_{0,\ell}$, where the subscript $0$ denotes isolated single spheres. In case of small spherical particles, when $ka \ll 1$, considering the dipole mode $\ell=1$, the spherical Bessel and Hankel functions can be approximated by their leading order terms: $j_{1}(k_{\text{in}}a) \sim k_{\text{in}}a/3$ and $h_{1}^{(2)}(k_{\text{out}}a) \sim i(k_{\text{out}}a)^{-2}$. Therefore the characteristic equation (\ref{charac_eqn_single_sphere}) reduces to $\epsilon_{\text{in}}=-2\epsilon_{\text{out}}$ which leads to the well known resonance frequency of $\omega=\omega_{p}/\sqrt{3}$ for an ideal electron gas with $\epsilon_{\infty}=1$ and $\gamma_s=0$.

\begin{figure}[h!]
\centering
\captionsetup{justification=centering}
\captionsetup[subfigure]{margin={0cm,0cm}}
\subfloat[]{
  \includegraphics[height=5cm,width=6.5cm]{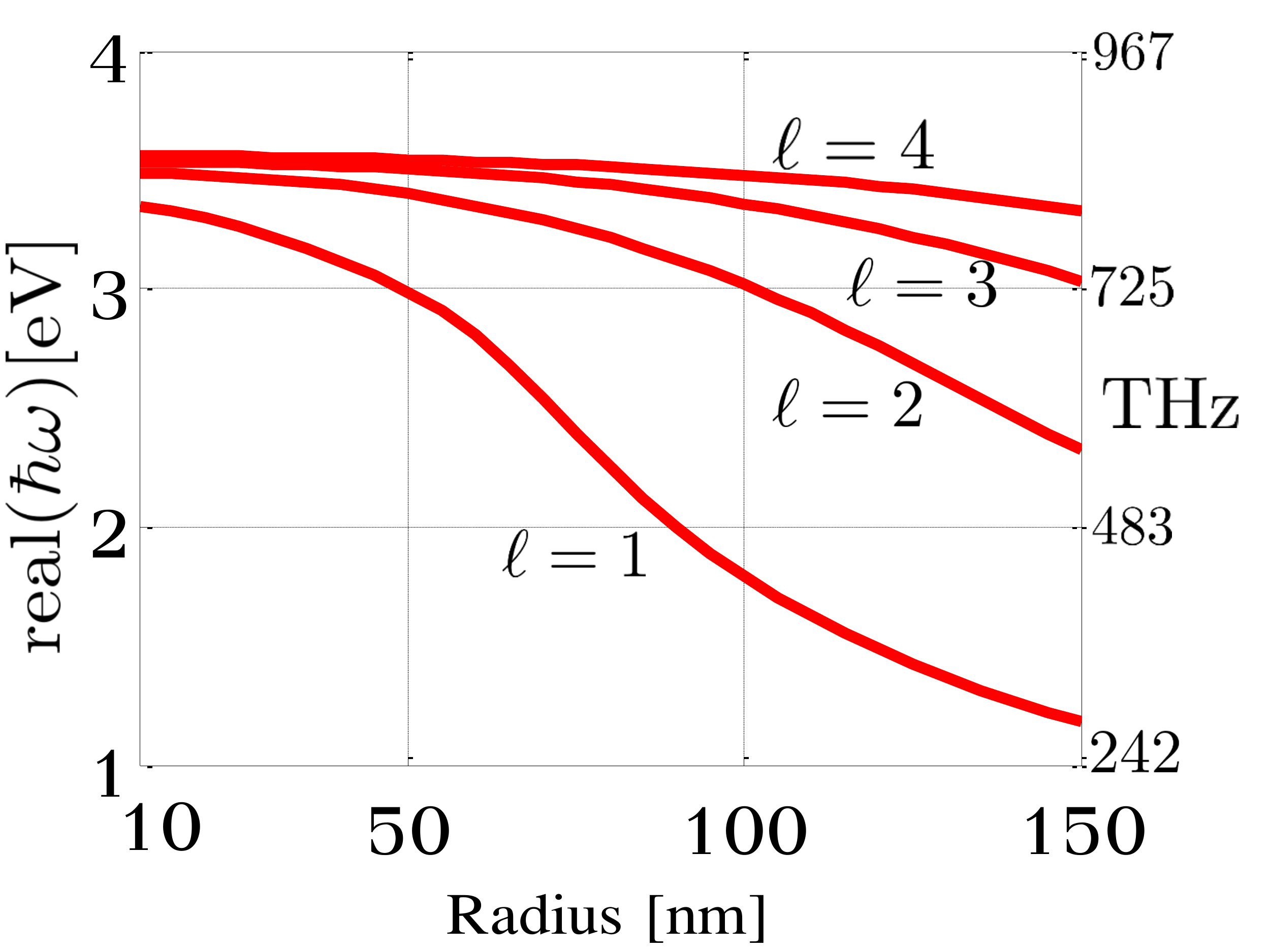}%
}

\captionsetup[subfigure]{margin={0cm,0cm}}
\hspace{-4mm}
\subfloat[]{%
  \includegraphics[height=5cm,width=7.0cm]{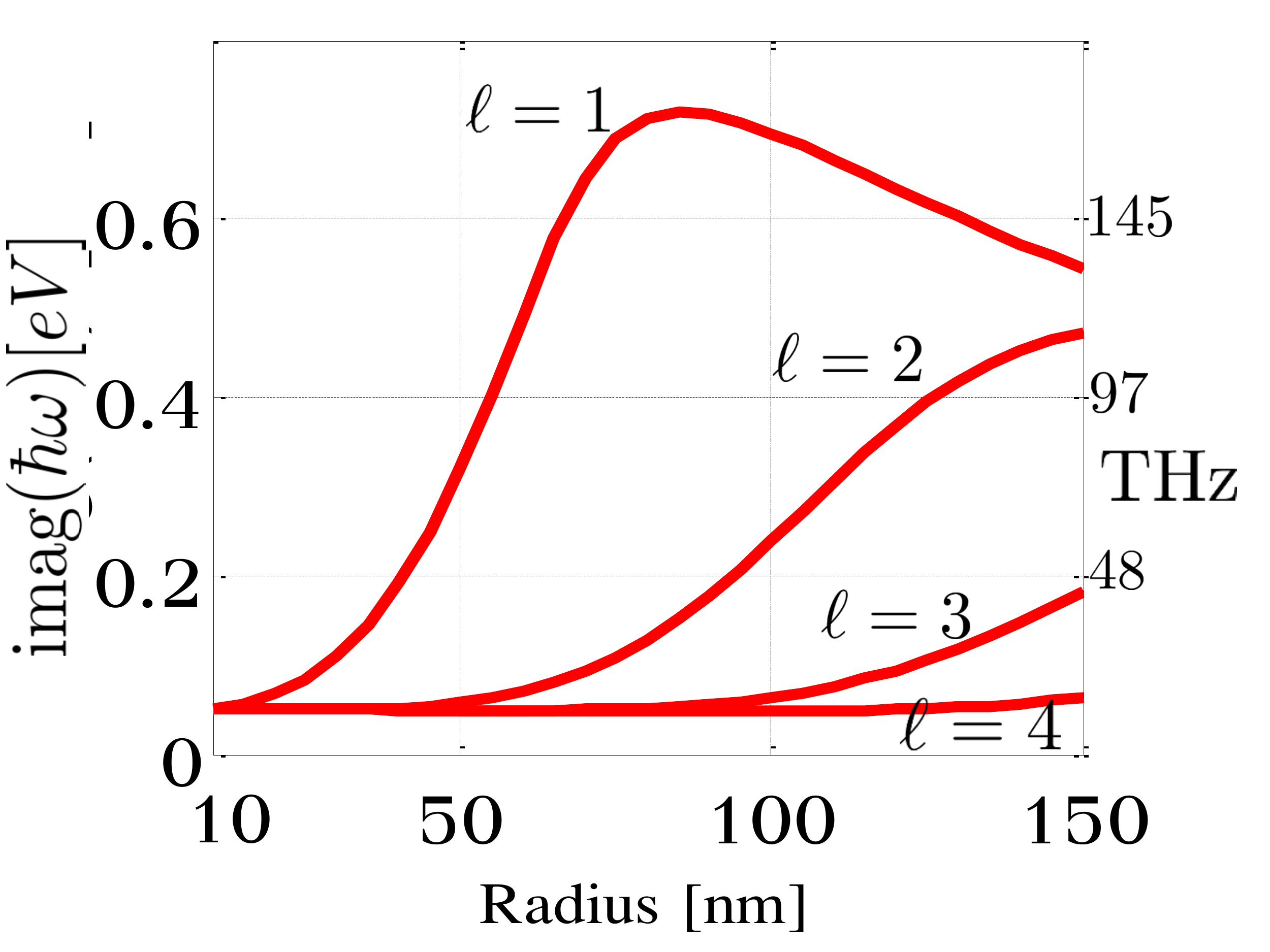}%
}

\captionsetup[subfigure]{justification=justified,singlelinecheck=false}
\caption{Eigenfrequencies for a metallic silver sphere with free space background as a function of radius for various $\ell$ values (a) real part and (b) imaginary part. The left and right axes represent values in units of eV and THz, respectively.} \label{SilverResonance}
\end{figure}

Next, we numerically solve eq. (\ref{charac_eqn_single_sphere}) for a silver sphere with a free space background. The parameters of the Drude-Sommerfeld model for silver are \cite{Silver_drude}: the plasma frequency $\omega_p=8.9$ eV, the damping rate $\gamma_s=0.1$ eV, and $\epsilon_{\infty}=5$. The eigenfrequencies are always complex, which indicates the radiative nature of the nanospheres \cite{SingleSphDAMPING}. The real and imaginary components of the eigenfrequencies as a function of radius and for various values of $\ell=1, 2, 3, 4$, are shown in Fig. \ref{SilverResonance}. The real part is the frequency required to excite a mode, for instance with a laser, and the imaginary component is the associated width of the mode. For all the modes, as expected, the real component decreases monotonically as the radius increases. To see the capability of the plasmonic sphere to manipulate light in sub-wavelength dimensions consider the dipole resonance ($\ell=1$) for a 50 nm sphere. The resonant frequency is about 3 eV corresponding to a free space wavelength of approximately 413 nm which is an order of magnitude larger than the radius of the sphere. The imaginary part of the eigenfrequencies consists of both non-radiative and radiative components,  $\text{Im}(\omega_{0,\ell})=\gamma^{\text{nrad}}+\gamma_{\ell}^{\text{rad}}$. The non-radiative damping is associated with the loss within the plasmonic sphere. It was discussed in \cite{SingleSphDAMPING} that the non-radiative component can approximately be considered size-independent and is of equal value for all different modes, $\gamma^{\text{nrad}}=(1/2) \gamma_{s}$. It is therefore clear from Fig. \ref{SilverResonance}(b) that the dipole is the most radiative mode. For $\ell=1$, initially the sphere becomes more radiative as the radius increases. However, larger spheres have less pronounced radiation properties. For all higher order modes, the imaginary part grows as the radius increases. 

\begin{figure}[h!]
\centering
\captionsetup{justification=centering}
\captionsetup[subfigure]{margin={0.3cm,0cm}}
\subfloat[]{
  \includegraphics[height=5.75cm,width=6.5cm]{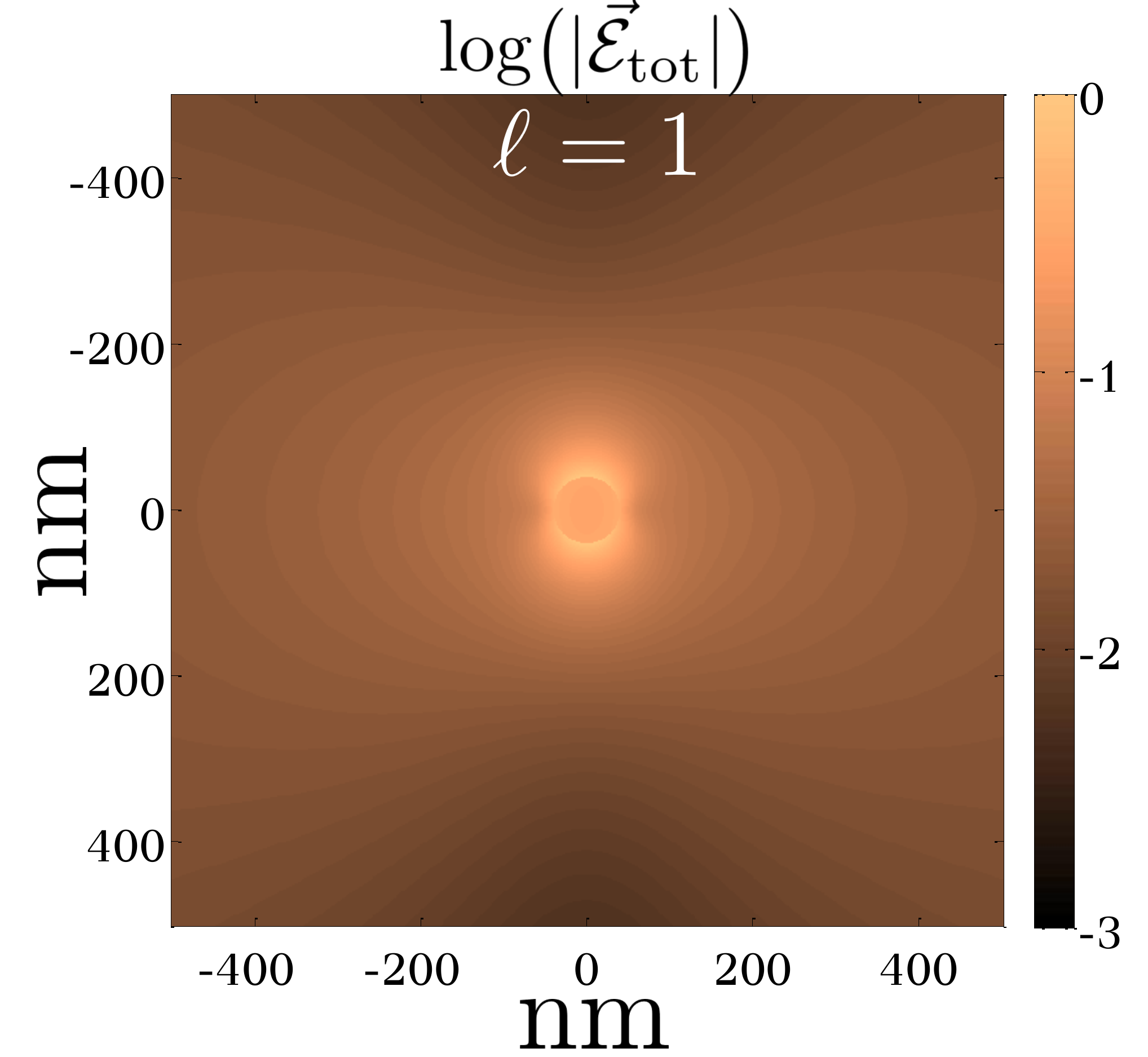}%
}
\captionsetup[subfigure]{margin={0.3cm,0cm}}
\hspace{-4mm}
\vspace{2mm}
\subfloat[]{%
  \includegraphics[height=5.75cm,width=6.5cm]{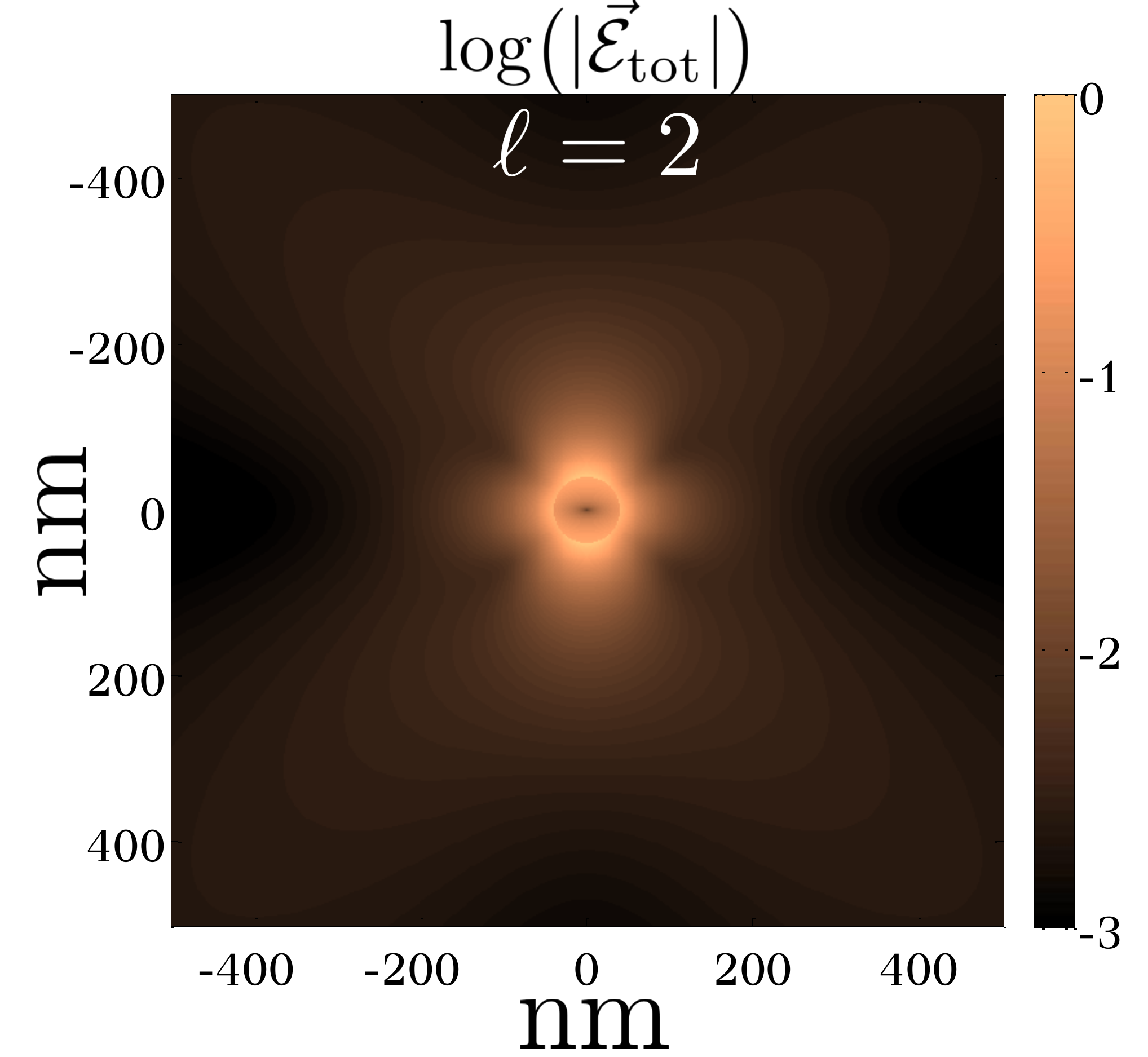}%
}
\captionsetup[subfigure]{margin={0.3cm,0cm}}
\hspace{-4mm}
\vspace{2mm}
\subfloat[]{%
  \includegraphics[height=5.75cm,width=6.5cm]{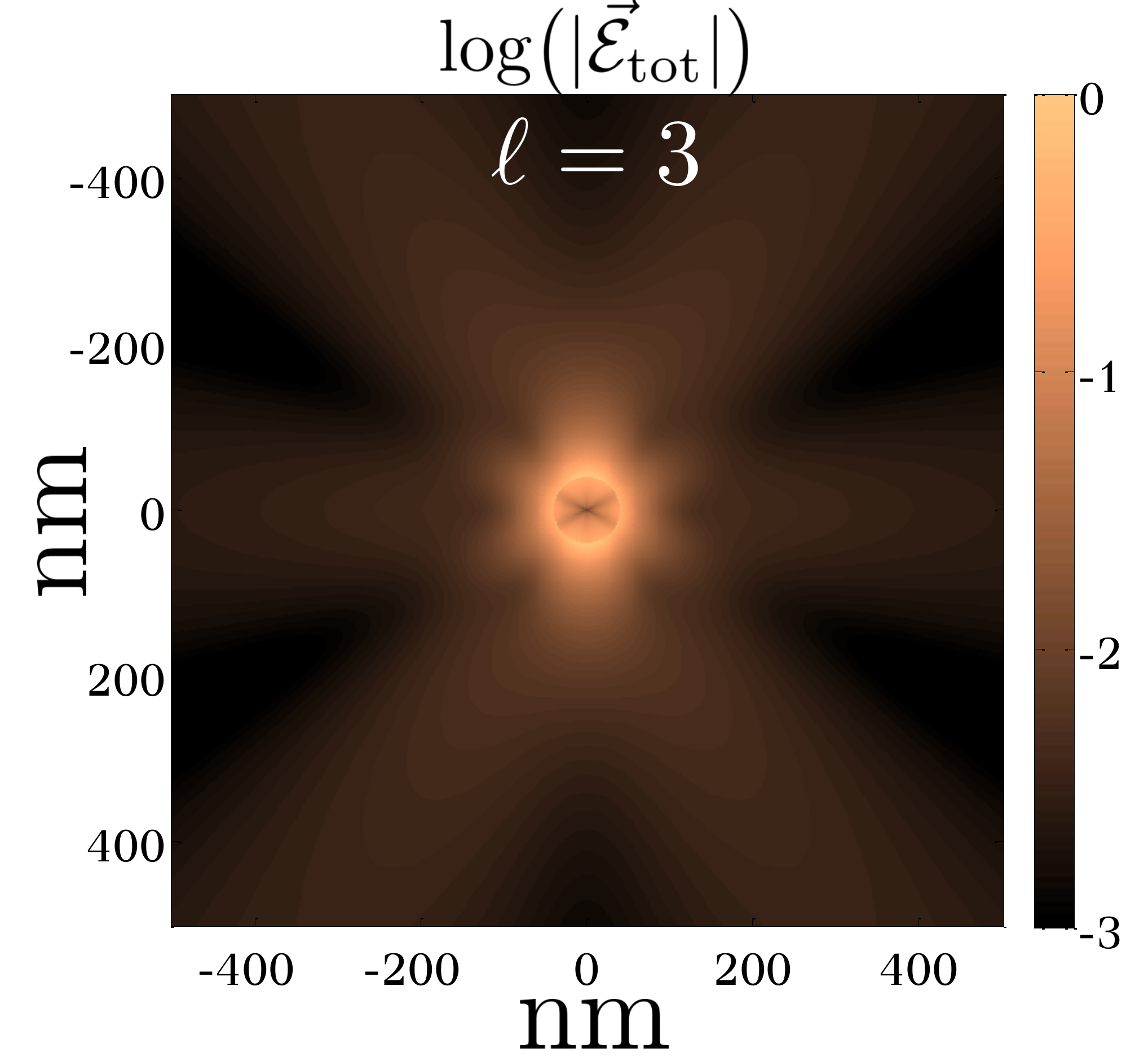}%
}
\captionsetup[subfigure]{justification=justified,singlelinecheck=false}
\caption{Electric field patterns of a silver sphere with a radius of 40 nm for (a) dipole mode, $\ell=1$, (b) quadrupole mode, $\ell=2$, and (c) octupole mode, $\ell=3$. The values are normalized to the maximum of the electric field. The projection $m$ is equal to zero for all three plots.} \label{Fieldplots}
\end{figure}

The electric field patterns of a silver sphere with a radius of 40 nm and various values of $\ell$ are shown in Fig. \ref{Fieldplots}. The field values are displayed logarithmically, and normalized to the maximum of the electric field. As the $\ell$ value increases the fields become more tightly bound to the surface of the sphere. This becomes important when we consider the coupling between two spheres in the next section.  

\begin{figure}[h!]
\captionsetup{justification=centering}
\centering
\captionsetup[subfigure]{margin={0.55cm,0cm}}
\subfloat[]{
  \includegraphics[height=3.5cm,width=3.5cm]{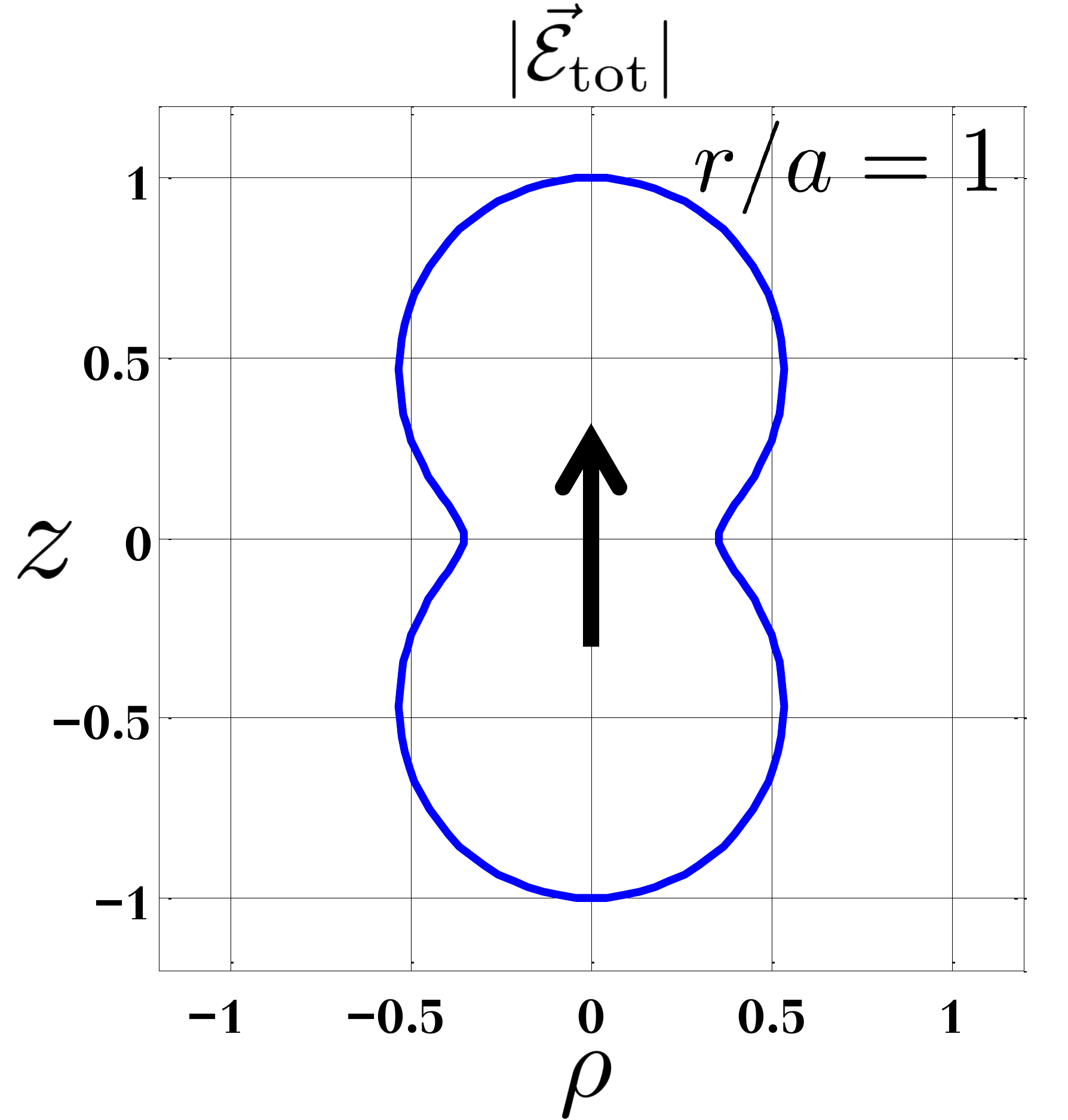}%
}
\vspace{1mm}
\captionsetup[subfigure]{margin={0.4cm,0cm}}
\hspace{1mm}
\subfloat[]{%
  \includegraphics[height=3.5cm,width=3.5cm]{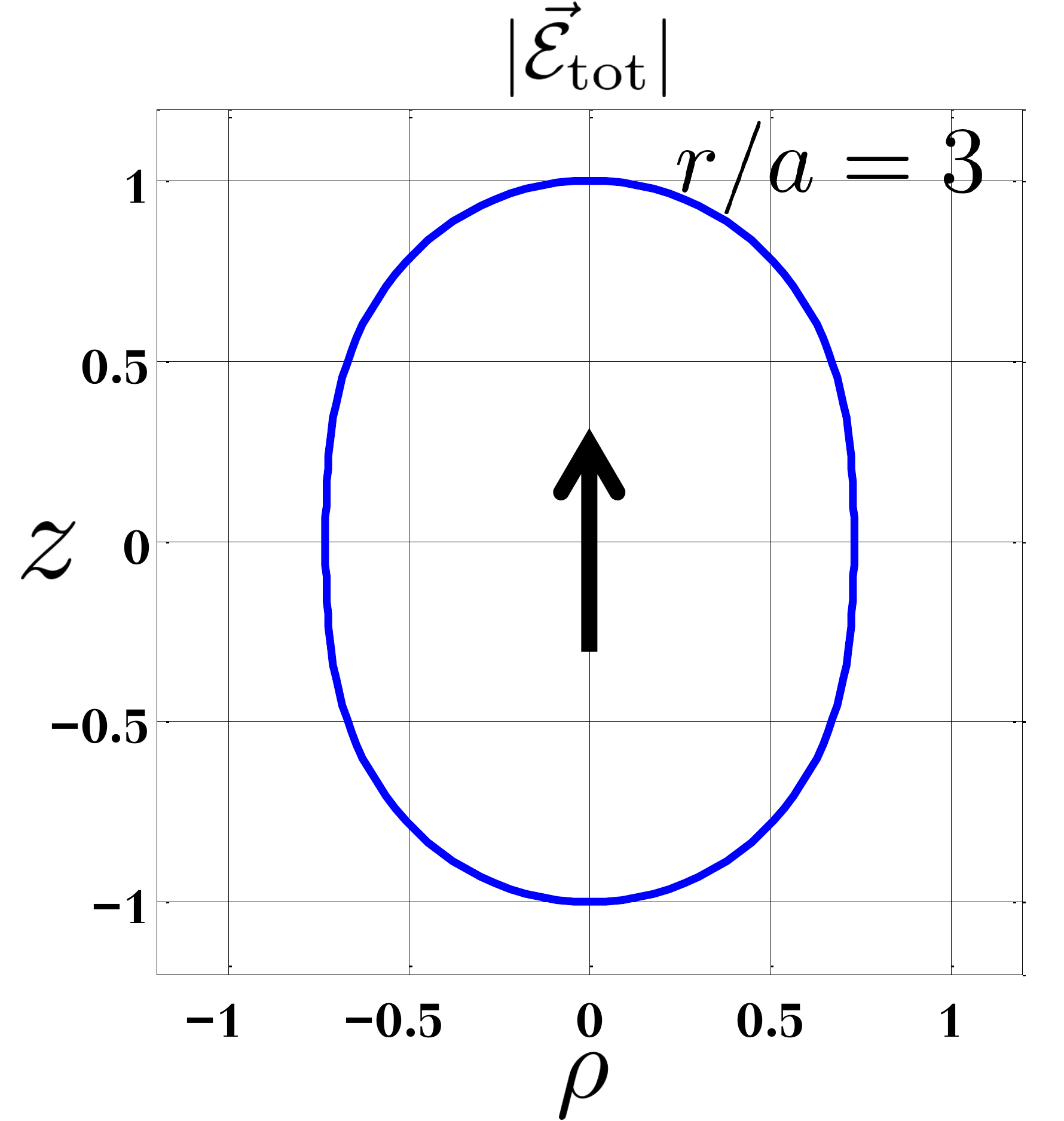}%
}
\vspace{1mm}
\captionsetup[subfigure]{margin={0.4cm,0cm}}
\hspace{1mm}
\subfloat[]{%
  \includegraphics[height=3.5cm,width=3.5cm]{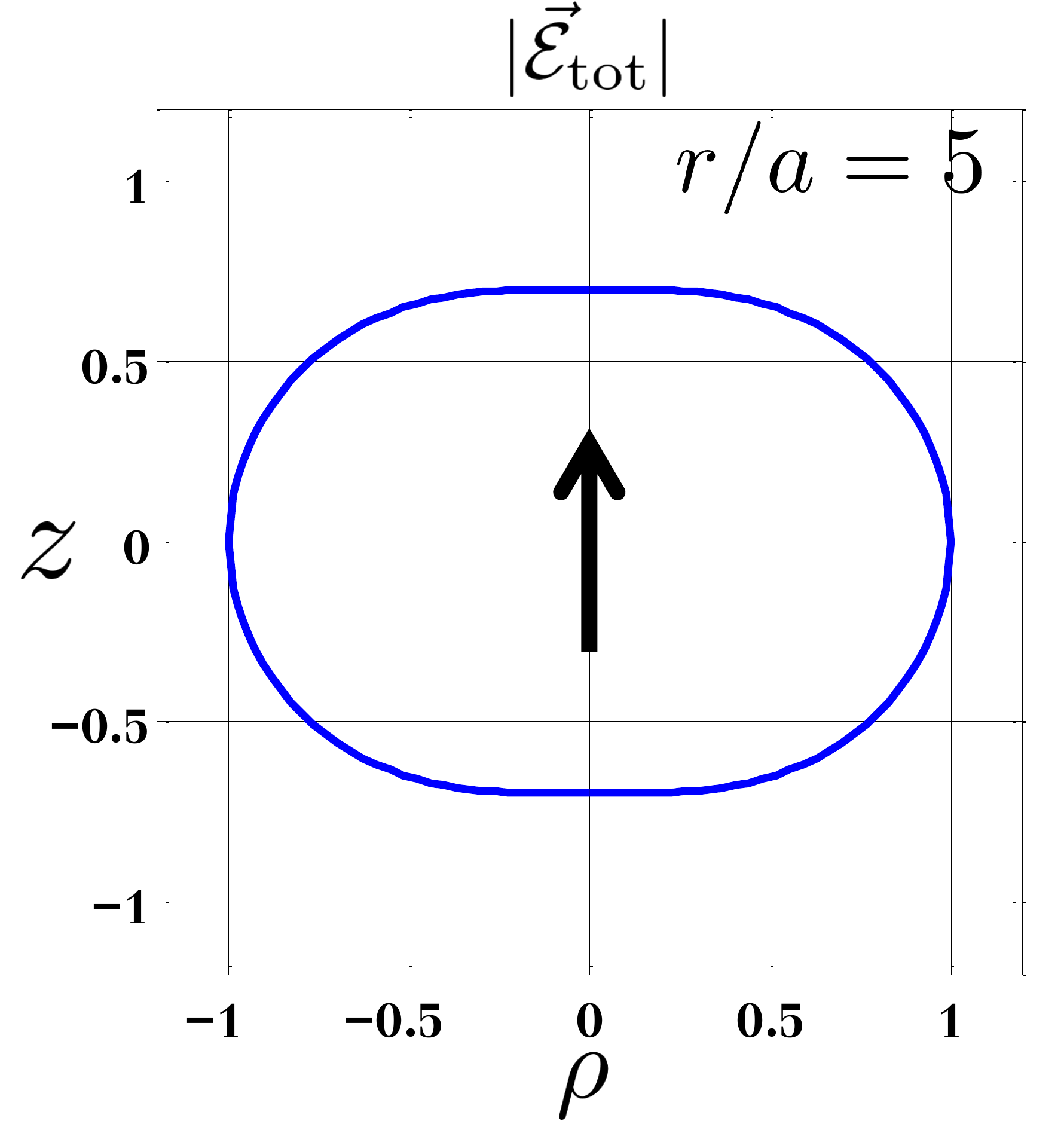}%
}
\vspace{1mm}
\captionsetup[subfigure]{margin={0.4cm,0cm}}
\hspace{1mm}
\subfloat[]{%
  \includegraphics[height=3.5cm,width=3.5cm]{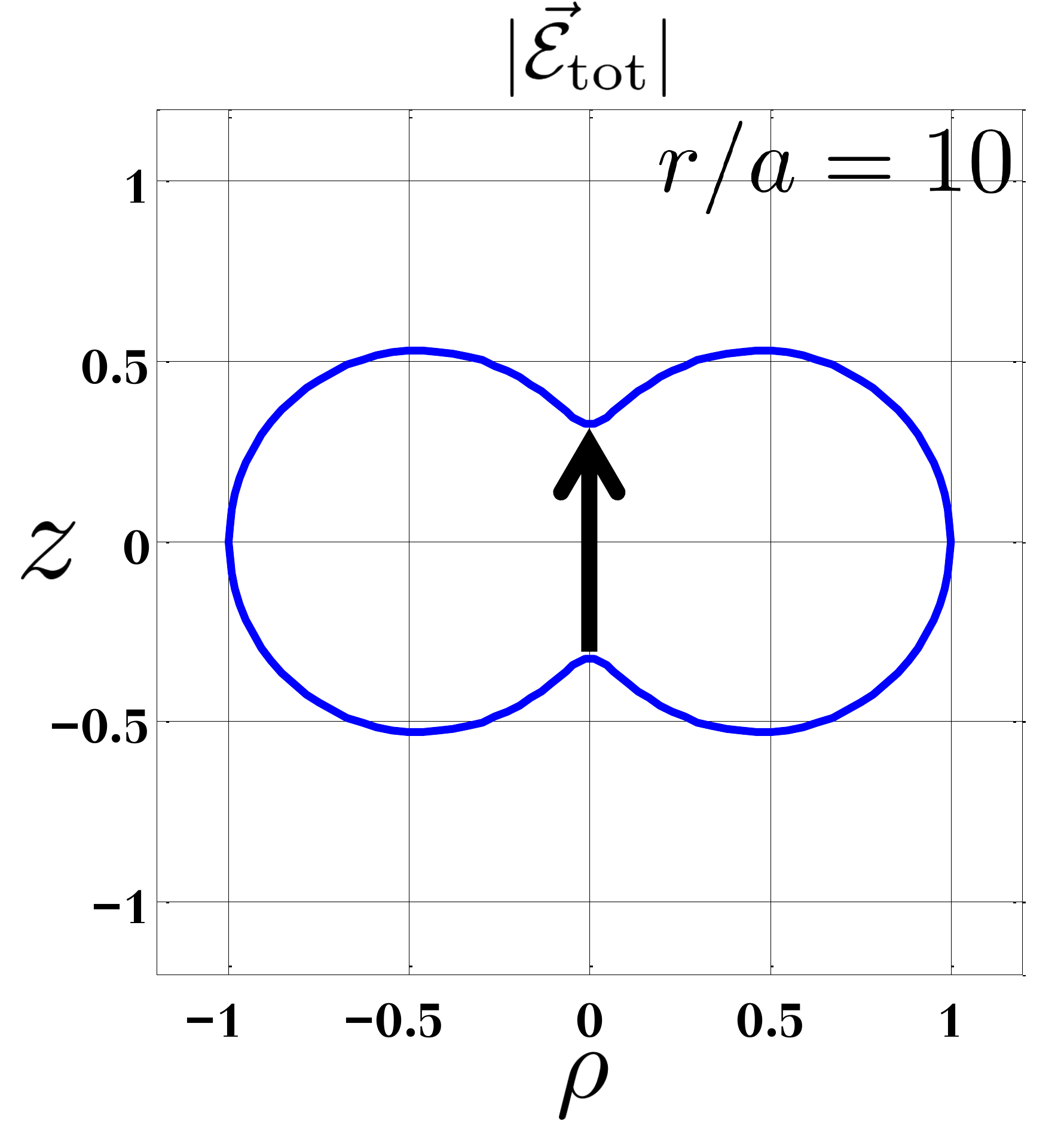}%
}
\vspace{1mm}
\captionsetup[subfigure]{margin={0.4cm,0cm}}
\hspace{1mm}
\subfloat[]{%
  \includegraphics[height=3.5cm,width=3.5cm]{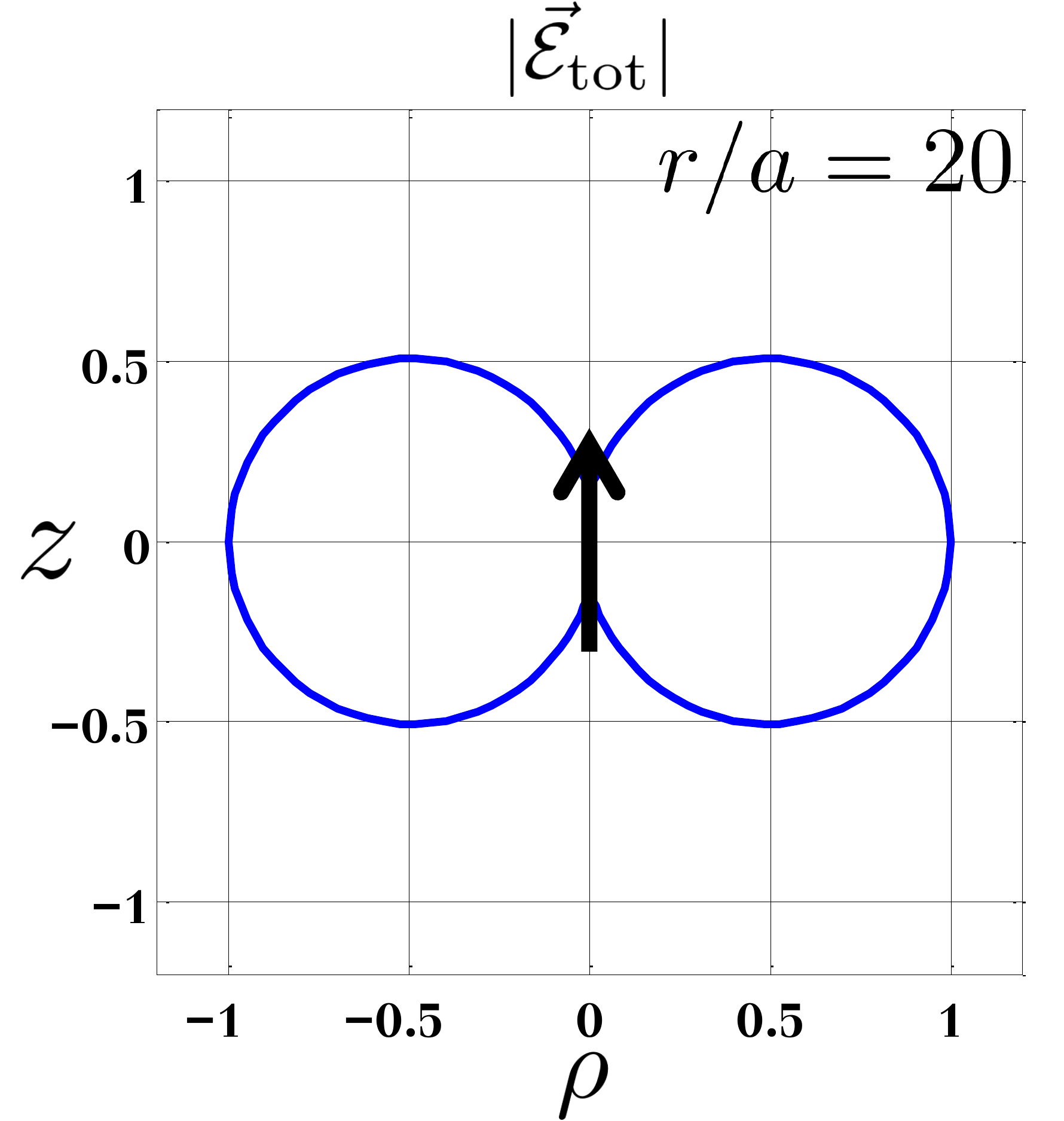}%
}
\vspace{2mm}
\captionsetup[subfigure]{margin={0.4cm,0cm}}
\hspace{1mm}
\subfloat[]{%
  \includegraphics[height=3.5cm,width=3.5cm]{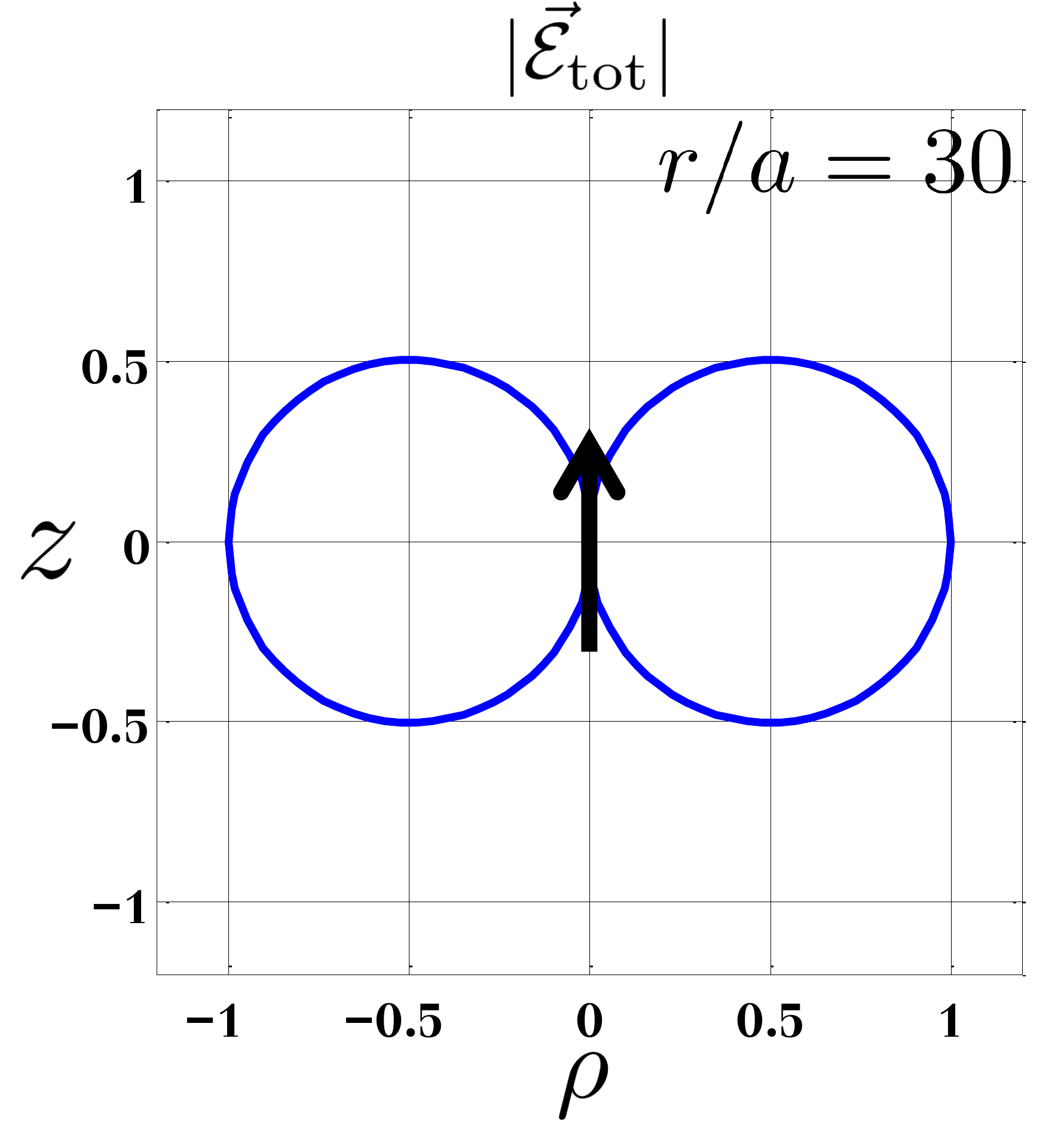}%
}
\captionsetup[subfigure]{justification=justified,singlelinecheck=false}
\caption{Electric field pattern for (a) $r/a=1$, (b) $r/a=3$, (c) $r/a=5$, (d) $r/a=10$, (e) $r/a=20$ and (f) $r/a=30$. $\rho$ is the radial distance and $z$ is the direction along the dipole. The values are normalized with respect to the maximum of the electric field.} \label{Field2Dplots}
\end{figure}

It is also instructive to look at the electric field pattern for the dipole mode. Fig. \ref{Field2Dplots} represents the electric field pattern in polar coordinates for various radial distances from the center of the sphere. The black arrow represents the dipole orientation. In all six figures, $\rho$ is the radial distance, perpendicular to the dipole axis and $z$ is the direction along the dipole. The blue line represents the relative field strength at a given polar angle. Furthermore, the field strength is normalized to the maximum value of the electric field. Fig. \ref{Field2Dplots}(a) shows the pattern at the surface of the plasmonic sphere, $r/a=1$. At locations close to the surface of the sphere, the fields reach a maximum along the direction of the dipole. At $r/a=3$ and $r/a=5$ the patterns are almost omnidirectional (see Figs. \ref{Field2Dplots}(b) and (c)). The well known torus shape radiation pattern of the dipole only emerges in the far field. This is shown in Figs. \ref{Field2Dplots}(d), (e) and (f).

The eigenmodes and plasmonic resonances of single metallic nanospheres of various sizes and the effect of different background dielectrics have been considered in detail in \cite{SingleSph3} for gold nanoparticles and in \cite{SingleSph2} for alkali metals such as sodium, lithium and Cesium; see \cite{SingleSphDAMPING} for a more detail description of the size dependency properties of nanospheres. Below we consider the case of two identical plasmonic spheres and discuss the coupling between them and the eigenmodes of the system.

\section{Superradiant and Dark States in The System of Two Coupled Spheres} \label{secIV}

Given the importance of two level systems in physics \cite{Amin_Holstein}, we now consider the case of two coupled metallic spheres within the coupled mode theory framework discussed earlier. The interaction between the spheres can greatly alter the radiation properties of the system and result in resonance frequencies profoundly different from those for the isolated spheres. Here, we limit our consideration to the dipole modes only. This is justified by the earlier discussion that the dipole mode is the most radiative mode and also has the longest range compared to higher multipolarities. The isolated frequencies discussed in the previous section serve as the diagonal elements in the coupled mode matrix (\ref{CMT_to_eff_Hamiltonian}). The real and imaginary components of the dipole eigenmode are the diagonal elements of the Hermitian (\ref{CMT_Matrix_hermitian}) and the anti-Hermitian (\ref{CMT_Matrix_nonhermitian}) matrices, respectively. The coupling between two modes, $\kappa$, makes up the off-diagonal matrix elements of the final matrix (\ref{CMT_to_eff_Hamiltonian}). 

However, a difficulty arises due to the normalization of the single sphere modes. In the far field region, $k_{\text{out}}r \gg 1$, the spherical Hankel functions behave asymptotically as
\begin{equation} \label{asymptotic_fields}
h_{\ell}^{(2)}(k_{\text{out}}r) \approx (i)^{\ell+1} \frac{e^{-ik_{\text{out}}r}}{k_{\text{out}}r}.
\end{equation} 
Because of the complex nature of the eigenfrequencies, the asymptotic form of the fields given by (\ref{asymptotic_fields}) grows exponentially in space as $r \rightarrow \infty$. This growth is however compensated by the exponential decay in time in the complete expression of the field (\ref{Phase_convention}) when the time dependency is considered. As a result the amplitude of the wave front of the total field (\ref{Phase_convention}) reaching any point in the asymptotic region  is proportional to $1/r$, as it is expected. Nevertheless, the modes (\ref{field_exprs}) should be properly normalized. The correct normalization of such modes was discussed in \cite{normalization_1D2,normalization_1D1,normalization_1D3,Amin_Thesis} for one-dimensional problems. The generalization to three dimensions by three different methods is discussed in \cite{normalization_3D1},  \cite{normalization_3D2}, and \cite{normalization_3D3}. However, in \cite{normalization_comparison}, it was shown that all three expressions are compatible. 

The normalization condition is given by \cite{normalization_comparison}:
\begin{equation} \label{normalization_int}
\int_V \sigma(\vec{r},\omega) \vec{\mathcal{E}}(\vec{r}).\vec{\mathcal{E}}(\vec{r}) d^3r + \frac{i\epsilon_{\text{out}}}{2k} \int_{\partial V} \vec{\mathcal{E}}(\vec{r}).\vec{\mathcal{E}}(\vec{r}) d^2r=1
\end{equation}
where $V$ is the integration volume and $\partial V$ is its surface. The integration volume is assumed to be sufficiently large, so the fields at its surface are accurately approximated by asymptotic expressions of the spherical Hankel function provided in (\ref{asymptotic_fields}). The modified dielectric function $\sigma(\vec{r},\omega)$ which incorporates the dispersiveness of the medium is given, according to \cite{DispersiveMedia} as
\begin{equation} \label{modified_dielectric}
\sigma(\vec{r},\omega)=\frac{1}{2\omega} \frac{\partial}{\partial \omega} \Big(\omega^2 \epsilon(\vec{r},\omega)\Big).
\end{equation}
Contrary to the normalization discussed earlier, the dot-product in (\ref{normalization_int}) does not require any complex conjugation of the fields. It is therefore easier to use the so-called tesseral harmonics instead of the conventional spherical harmonics in the field expression (\ref{field_exprs}). The tesseral harmonics (sometimes also called real spherical harmonics) are nothing but even and odd superpositions of the traditional spherical harmonics, see Appendix \ref{THandIs}. The normalization condition (\ref{normalization_int}) defines the constant $\zeta$ in the field expressions (\ref{field_exprs}) up to a phase. It is shown in Appendix \ref{QNMNAp} that assuming the volume of integration as a sphere, the volume and surface terms in the normalization expression can be evaluated explicitly. It is furthermore proved in the same appendix that the condition (\ref{normalization_int}) reduces to:
\begin{equation} 
I \big[ j_{\ell}(k_{\text{in}}a) \big] -  I \big[h_{\ell}^{(2)}(k_{\text{out}}a)\big]=1,
\end{equation}
where the functional $I \big[f_{\ell}(kr) \big]$ is given by (\ref{volumeterm4})
\begin{widetext}
\begin{align} 
I \big[f_{\ell}(kr)\big]=\sigma(\vec{r},\omega) \ \zeta^2 \ C^2(r;a)\frac{\ell(\ell+1)}{k^2} \bigg[ r f_{\ell}^2(kr)+kr^2 f_{\ell}(kr) f_{\ell}^{'}(kr) 
+\frac{k^2r^3}{2}\Big(f_{\ell}^2(kr)-f_{\ell-1}(kr)f_{\ell+1}(kr) \Big) \bigg].
\end{align}
\end{widetext}

Once the normalization constant $\zeta$ is found the coupling between two modes can be calculated according to 
\begin{equation} \label{coupling_coefficient_modified}
\kappa=-\frac{1}{2}\omega_0 \int_{\text{V}_1}  \big(\epsilon_1(\vec{r})-1 \big) \vec{\mathcal{E}}^{1}(\vec{r}). \vec{\mathcal{E}}^2(\vec{r}) d^3r,
\end{equation} 
where $\text{V}_1$ is again the volume of sphere 1. Contrary to (\ref{coupling_coefficient}), the fields from both spheres are treated on equal footing and no complex conjugation is required due to the normalization definition (\ref{normalization_int}). 

In what follows we show the eigenfrequencies of a system of two coupled spheres with different sizes and different separation distances. The coupling $\kappa$ is calculated numerically and the coupled mode matrix (\ref{CMT_to_eff_Hamiltonian}) is constructed and diagonalized. In the first case two identical silver spheres with radii of 10 nm are considered and the eigenfrequencies of the system are calculated for two different dipole orientations and as a function of the separation distance between the spheres. It is important to note that due to the symmetry of the dipole modes, dipoles with perpendicular orientations do not couple. Therefore we consider two parallel orientations only. 

As a check of the coupled mode theory approximation, we have also computed the resonant frequencies of two coupled metallic spheres via a modal solution. We refer to this modal solution as exact even though we only retain a finite number of modes that adequately resolves the resonant frequencies. The modal solutions are rigorously based upon solving Maxwell’s equations, and as such we treat them as exact, in contrast to the approximations used to generate the coupled mode theory results. 

The modal solution is derived by considering the discrete modes for an isolated sphere, which have electric fields as given by (\ref{field_exprs}), and corresponding magnetic fields determined via Maxwell’s equations. If we numerically truncate the modes at a maximum $\ell$ value, then this results in a finite set of possible $\ell$ and $m$ values. Each of these ($\ell$,$m$) combinations can be considered as defining a basis function, with distinct basis functions defined for the region interior and exterior to the sphere. The weighted summation of these basis functions can then approximate the fields supported everywhere by the metallic sphere. 

We now consider two spheres with these allowed modal solutions, and our goal is to find which particular combination of modal coefficients satisfies the boundary condition of the tangential electric and magnetic fields being continuous at the surfaces of both spheres. This is accomplished by choosing $N$ points spaced approximately equally over the surface of each sphere, where $N$ exceeds the number $M$ of modal coefficients we wish to determine. 

Each point on the surface of a sphere defines a constraint equation, with $4N$ total constraint equations due to there being four field components to match at each surface point: two tangential electric field components, and two magnetic field components. If we were able to exactly represent the fields, then at resonance the sum of the mode contributions at each point would equal zero, resulting in a matrix equation $Ax=0$ involving a $4N$-by-$M$ matrix $A$ and a length-$M$ vector $x$ representing the mode coefficients.  

However, because we are working with a truncated set of modes, we cannot solve the boundary condition exactly. We instead seek complex frequencies at which the smallest singular value of $A$ achieves a local minimum. These frequencies are approximately equal to the true resonant frequencies, and for the results plotted in this paper, frequency convergence was observed for maximum $\ell$ values ranging from 2 to 8, with a larger $\ell$ value being required as the sphere separation decreases. 

\begin{figure}[h!]
\captionsetup{justification=centering}
\vspace{-4.5mm}
\centering
\captionsetup[subfigure]{margin={1.2cm,0cm}}
\hspace{-3mm}
\subfloat[]{
  \includegraphics[scale=0.45]{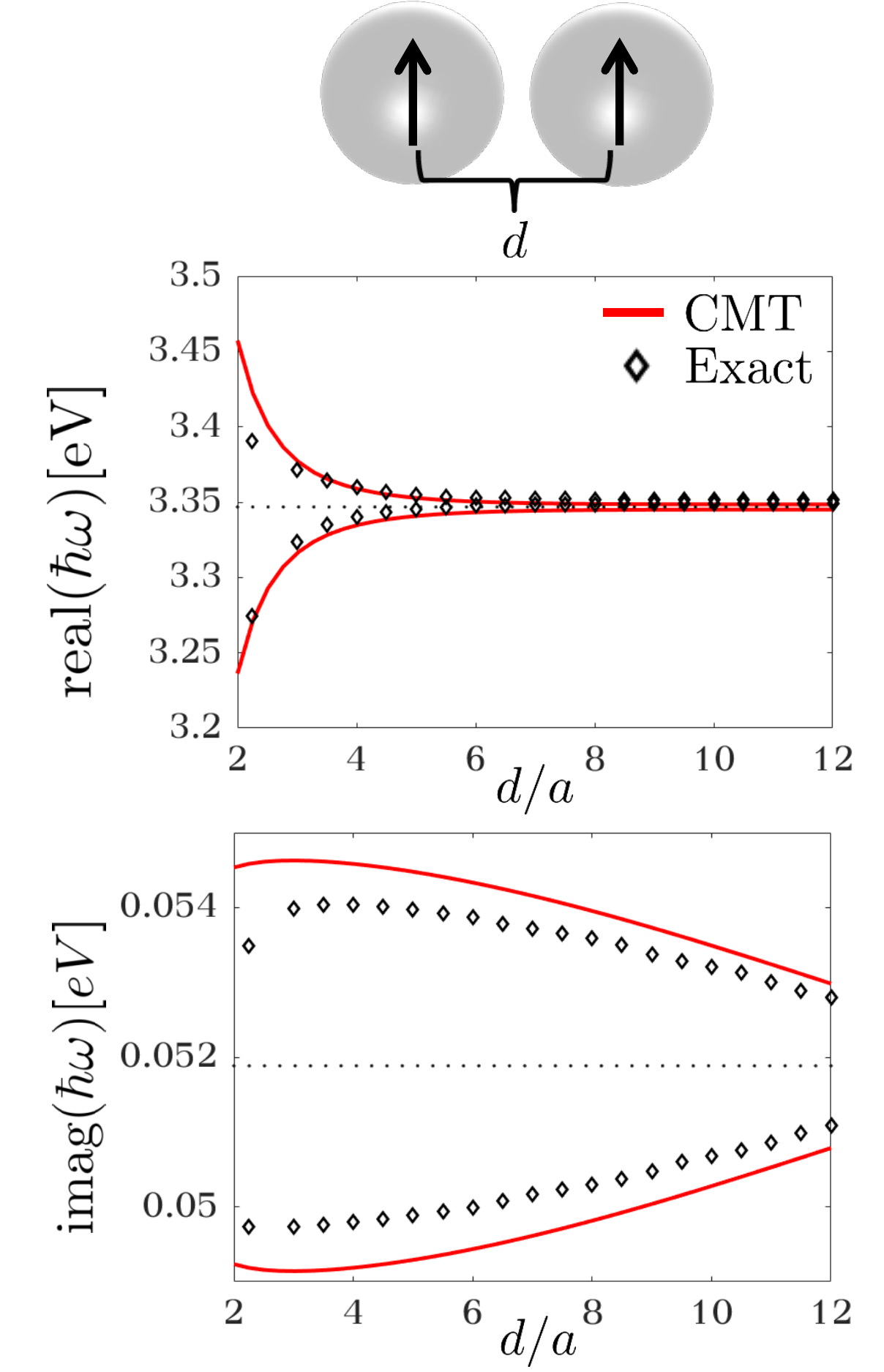}%
}   
\vspace{1mm}
\captionsetup[subfigure]{margin={1.2cm,0cm}}
\hspace{0mm}
\subfloat[]{%
  \includegraphics[scale=0.45]{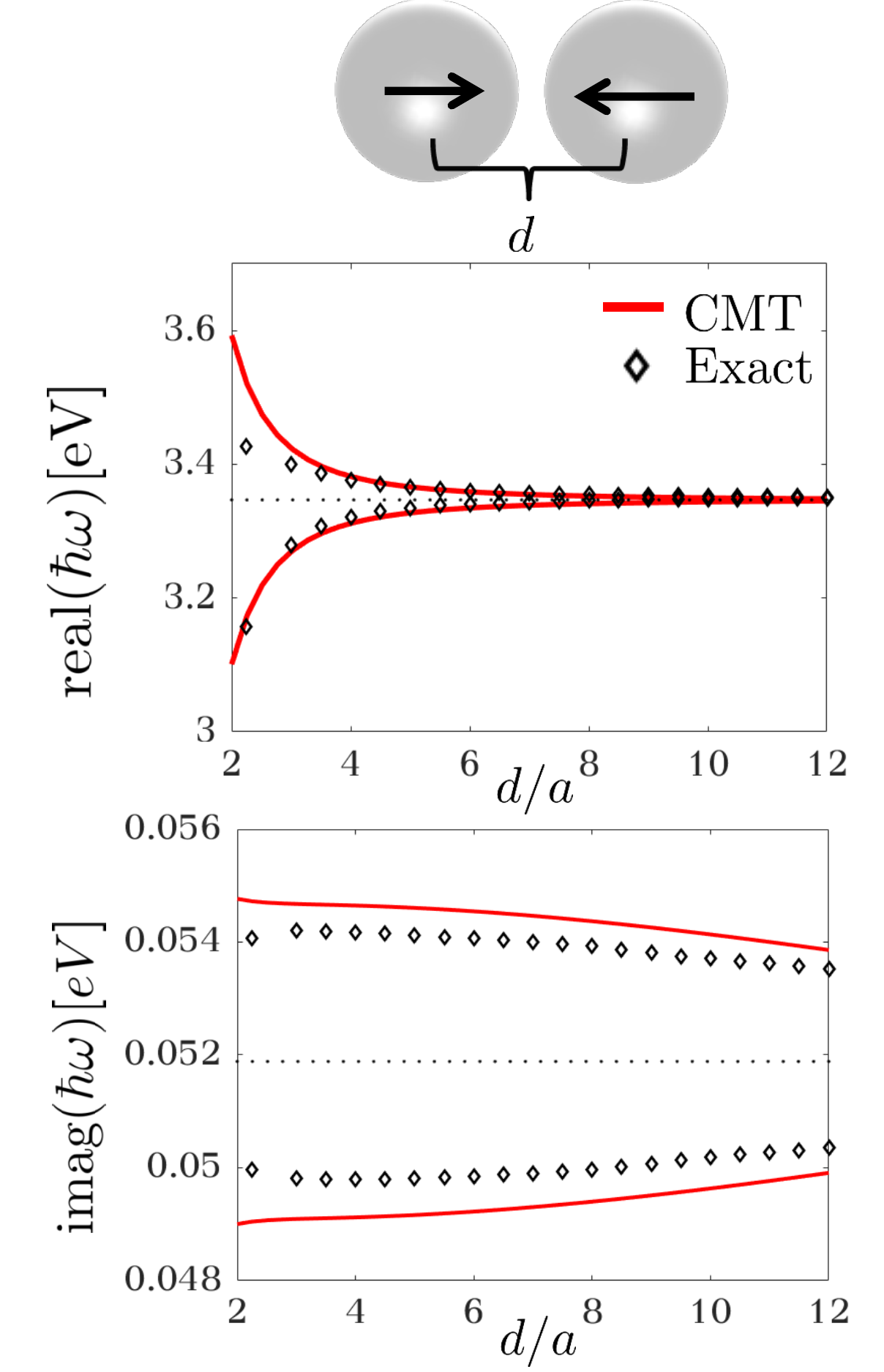}%
}
\captionsetup[subfigure]{justification=justified,singlelinecheck=false}
\caption{Real and imaginary components of the eigenfrequencies of a system of two coupled identical silver spheres with (a) vertical and (b) horizontal dipole orientations calculated via coupled mode theory (CMT) shown in red (solid lines) and modal expansion (Exact) shown in black (diamonds). The radii of the spheres are 10 nm. $d$ is the center-to-center separation and $a$ is the radius of the spheres. The black dotted lines represent the unperturbed eigenfrequency of a single sphere.} \label{coupling_10nm_sphs}
\vspace{5mm}
\end{figure}

Fig. \ref{coupling_10nm_sphs}(a) shows the case where the two dipoles are parallel (vertical orientation). Both the real and imaginary components of the two eigenfrequencies of the system are plotted as a function of the separation $d$ between the two spheres normalized by the sphere radius $a$. The red solid lines and the black diamonds correspond to calculations via coupled mode theory (CMT) and modal expansion (Exact), respectively. There is good agreement between the two methods across the entire separation distances, even for small separations or strong interaction regime. This indicates that the dipole mode has the largest contribution in the coupling strength between the spheres and therefore coupled mode theory provides an acceptable approximation of the solution. The maximum coupling occurs when the two spheres are in contact with one another which in turn results in the largest deviation from the unperturbed frequency of a single sphere (dotted black lines). As the separation increases, the coupling decreases and the two eigenfrequencies approach the unperturbed eigenmode. A similar phenomenon is seen for the case of two dipoles in Fig. \ref{coupling_10nm_sphs}(b) (horizontal orientation). However, for a small separation distance between the spheres, the splitting of the two eigenmodes is larger for horizontal (colinear) dipole orientation than for vertical (parallel) dipole orientation. This is because, as shown in the previous section, the electric field is maximum along the dipole direction in the near field (see Fig. \ref{Field2Dplots}). Consequently, this orientation results in larger coupling coefficients between the spheres. 

\begin{figure}[h!]
\vspace{10mm}
\centering
\captionsetup{justification=centering}
\includegraphics[scale=0.5]{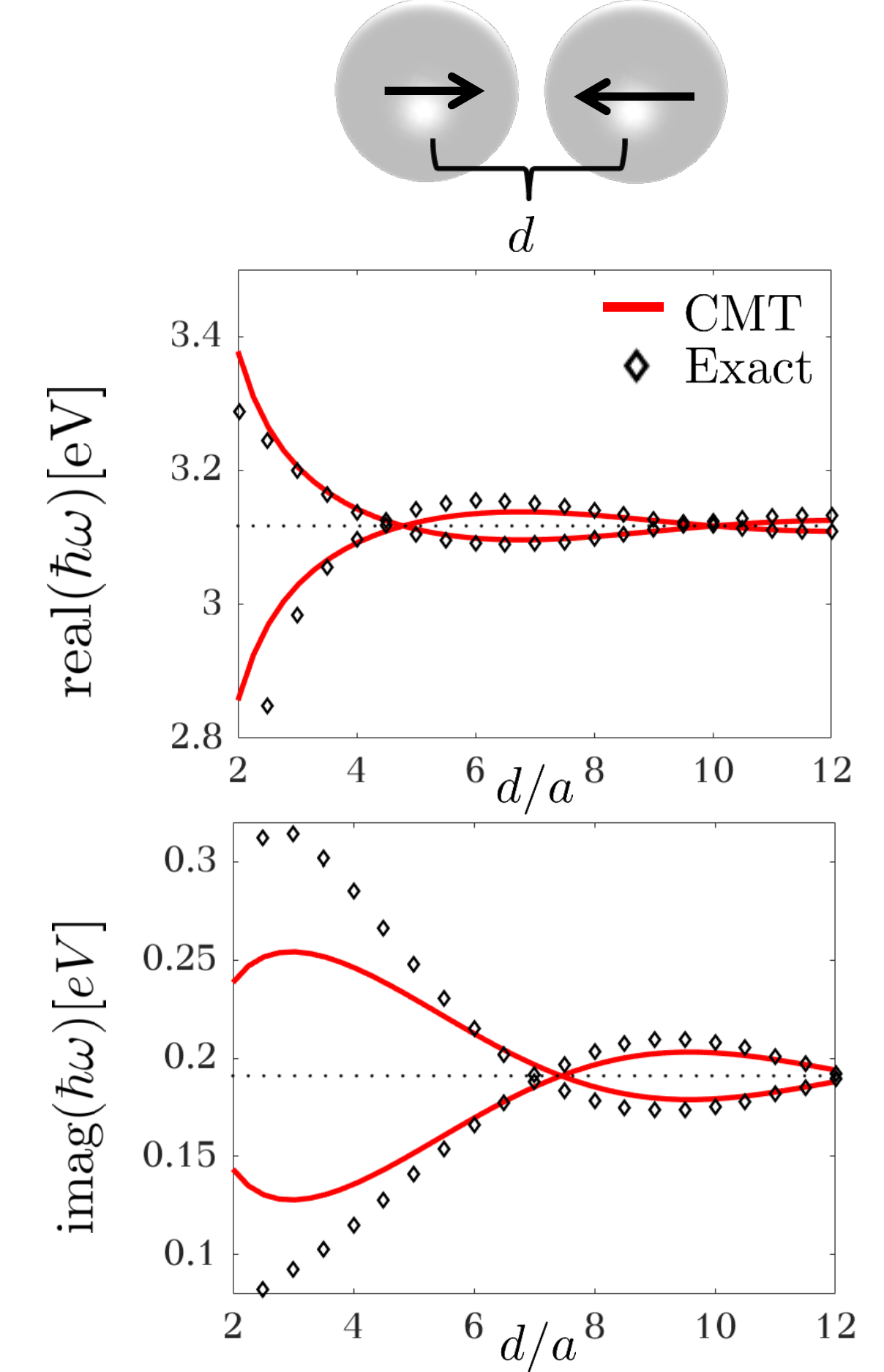}
\caption{Real and imaginary components of the eigenfrequencies of a system of two coupled identical silver spheres with horizontal dipole orientation calculated via coupled mode theory shown in red (solid lines) and modal expansion shown in black (diamonds). The radii of the spheres are 40 nm. $d$ is the center-to-center separation and $a$ is the radius of the spheres. The black dotted lines represent the unperturbed eigenfrequency of a single sphere.}
\label{coupling_40nm}
\vspace{10mm}
\end{figure}

According to our findings in the previous section, silver spheres with the larger 40 nm radii are more radiative than the 10 nm radii spheres just considered. It is therefore desirable to look at the case of coupling between larger spheres since the coupling is stronger. Fig. \ref{coupling_40nm} shows the eigenfrequencies of a system of two silver spheres with radii of 40 nm. Because the coupling coefficient between two horizontal dipoles is greater than that of two vertical dipoles, we only consider the horizontal case. The difference between the superradiant and subradiant states is more pronounced in this case. At $d/a \approx 3$, coupled mode theory predicts a maximal difference between the imaginary components of the two eigenmodes. Similar to the previous case, the two eigenfrequencies approach the unperturbed resonance as the separation distance increases. At lower separations however, when the spheres are strongly interacting, coupled mode theory solution deviates more from the exact solution compared to the 10 nm spheres shown in Fig. \ref{coupling_10nm_sphs}(b). This indicates the importance of higher order modes in the interaction strength of larger spheres, and that the coupled mode results can improve if these modes are included in the calculations.

\begin{figure}[h!]
\vspace{10mm}
\centering
\captionsetup{justification=centering}
\includegraphics[scale=0.5]{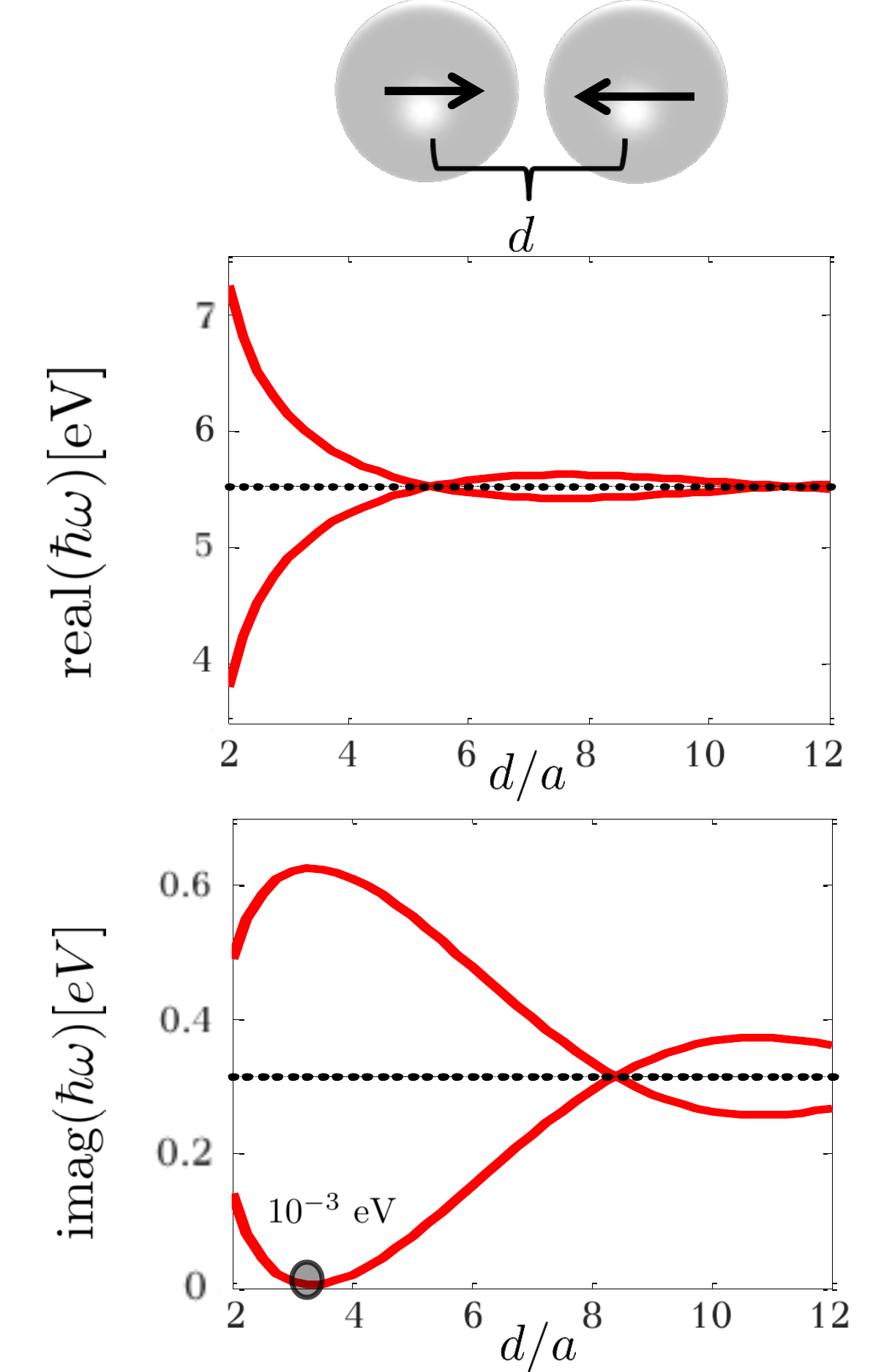}
\caption{Real and imaginary components of the eigenfrequencies of a system of two coupled identical spheres with horizontal dipole orientation. The material properties of the spheres are: $\epsilon_{\infty}=1$, $\omega_p=10.918$ eV and $\gamma_s=0$. The radii of the spheres are 20 nm. $d$ is the center to center separation and $a$ is the radii of the spheres. The black dotted lines represent the unperturbed eigenfrequency of a single sphere.}
\label{coupling_darkmode}
\vspace{10mm}
\end{figure}

According to our calculations, an exact dark mode does not exist for a system of two silver plasmonics dipoles. Therefore a numerical search over the parameters of the Drude-Sommerfeld dielectric function (\ref{Drude_dielectric_func}) was performed in order to find material properties for which two plasmonic spheres can support a dark mode. Fig. \ref{coupling_darkmode} shows the eigenfrequencies of a system of two spheres with Drude-Sommerfeld parameters of $\epsilon_{\infty}=1$, $\omega_p=10.918$ eV and $\gamma_s=0$. At $d/a \approx 3.5$, the rank of the anti-Hermitian part of the coupled mode matrix is almost unity indicating that the interaction between the two spheres occurs through a single continuum channel. Consequently, the imaginary component of the subradiant mode is extremely small, in the order of 10$^{-3}$ eV. This is indicated with a black circle in the figure. 

\section{Plasmonic Waveguide} \label{secV}

\begin{figure}[h!]
\vspace{10mm}
\centering
\captionsetup{justification=centering}
\includegraphics[scale=0.7]{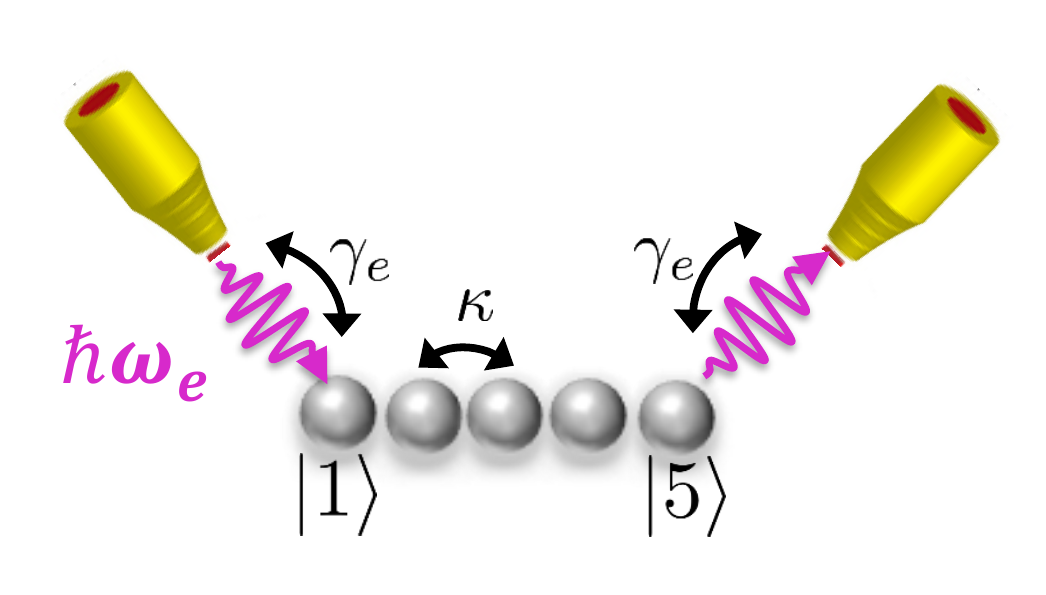}
\caption{Schematics of a plasmonic waveguide; a one-dimensional chain of five silver spheres with nearest neighbor coupling $\kappa$. The two edges are symmetrically coupled to continuum, the excitation source with frequency $\omega_e$, with coupling coefficient $\gamma_e$.}
\label{plasmonic_waveguide_schematics}
\vspace{10mm}
\end{figure}

We now consider the signal transmission through a plasmonic waveguide, namely a one-dimensional chain of identical spheres. The idea that such a structure acts as a waveguide due to interparticle coupling was proposed in \cite{Quinten98} and experimentally verified in \cite{waveguide_exp}. Here, it is assumed that the two edges of the waveguide are connected to an instrument capable of exciting the system of spheres with frequency $\omega_e$ and measuring the electric field intensity. Fig. \ref{plasmonic_waveguide_schematics} depicts the schematic of the plasmonic waveguide and the two probes symmetrically coupled to the edges of the chain with coupling constants $\gamma_e$.  Similar to the tight binding model of crystals in condensed matter physics, it is further assumed that each sphere in the chain only interacts with its nearest neighbor. This system can be modeled with the effective non-Hermitian Hamiltonian (\ref{ReducedHeff}) 
\begin{equation}   \label{effective_hamiltonian_plasmonic_waveguide}
\mathscr{H}_{\text{eff}}=
\begin{bmatrix}
    \frac{i}{2}\gamma_e+\omega_0 & \kappa & 0 & \dots & 0 & 0 \\
    \kappa & \omega_0 & \kappa & \dots & 0 & 0 \\
    0 & \kappa & \omega_0 & \kappa & \dots & 0  \\
    \vdots & \vdots & \vdots & \ddots & \vdots & \vdots \\
    0 & 0 & 0 & \dots & \kappa & \frac{i}{2}\gamma_e+\omega_0 
\end{bmatrix}
,
\end{equation}
where $\omega_0$ is the unperturbed dipole frequency of an isolated sphere, and $\kappa$ is the coupling coefficient (\ref{coupling_coefficient_modified}) between adjacent dipoles. It is important to mention that the addition of the anti-Hermitian matrix elements, $\frac{i}{2}\gamma_e$, with a positive sign is due to the phase convention adopted earlier in (\ref{Phase_convention}). 

\begin{figure}[h!]
\vspace{7mm}
\centering
\captionsetup{justification=centering}
\captionsetup[subfigure]{margin={0cm,0cm}}
\subfloat[]{
  \includegraphics[height=6cm,width=8cm]{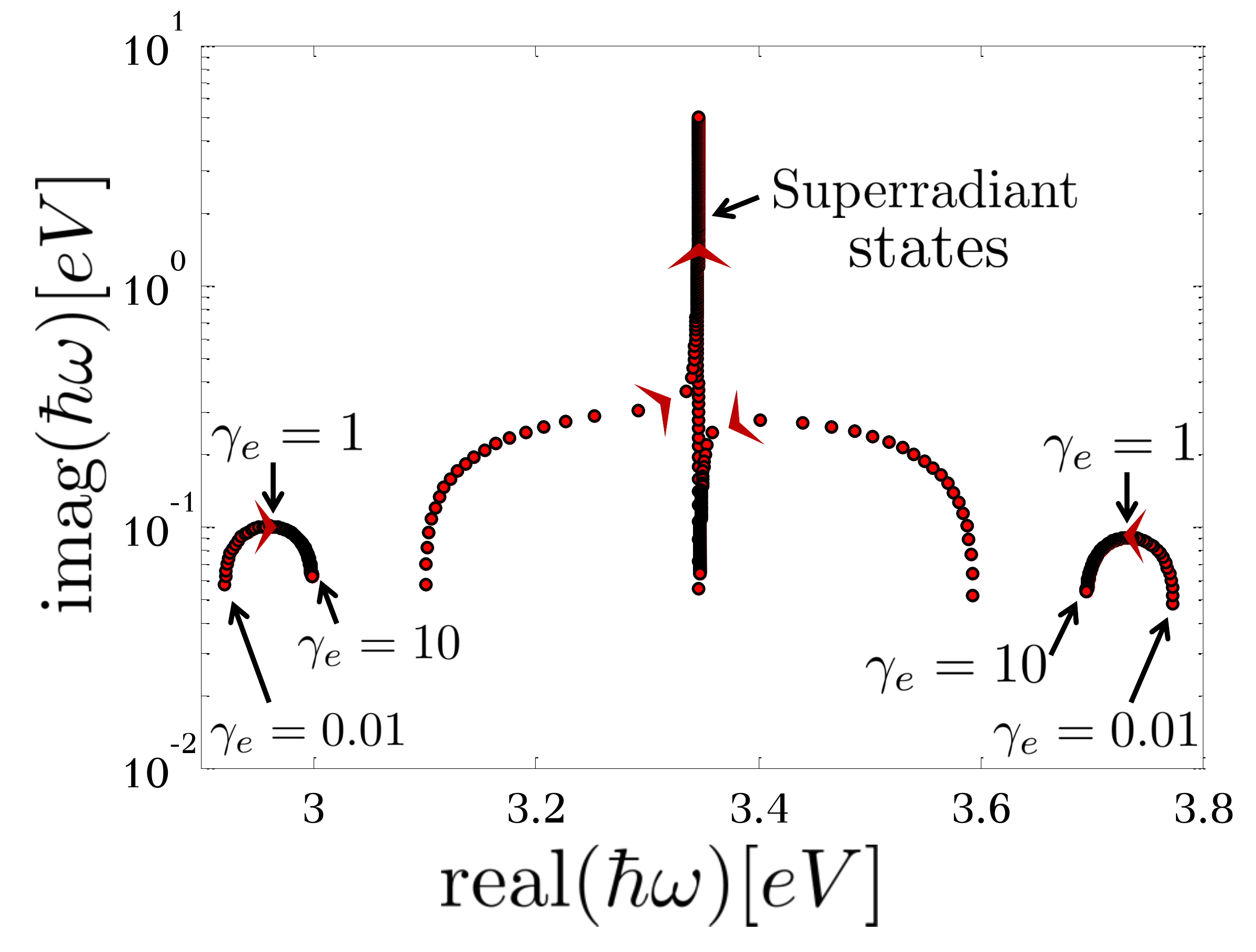}%
}

\captionsetup[subfigure]{margin={0cm,0cm}}
\hspace{-0.5mm}
\subfloat[]{%
  \includegraphics[height=6cm,width=8.2cm]{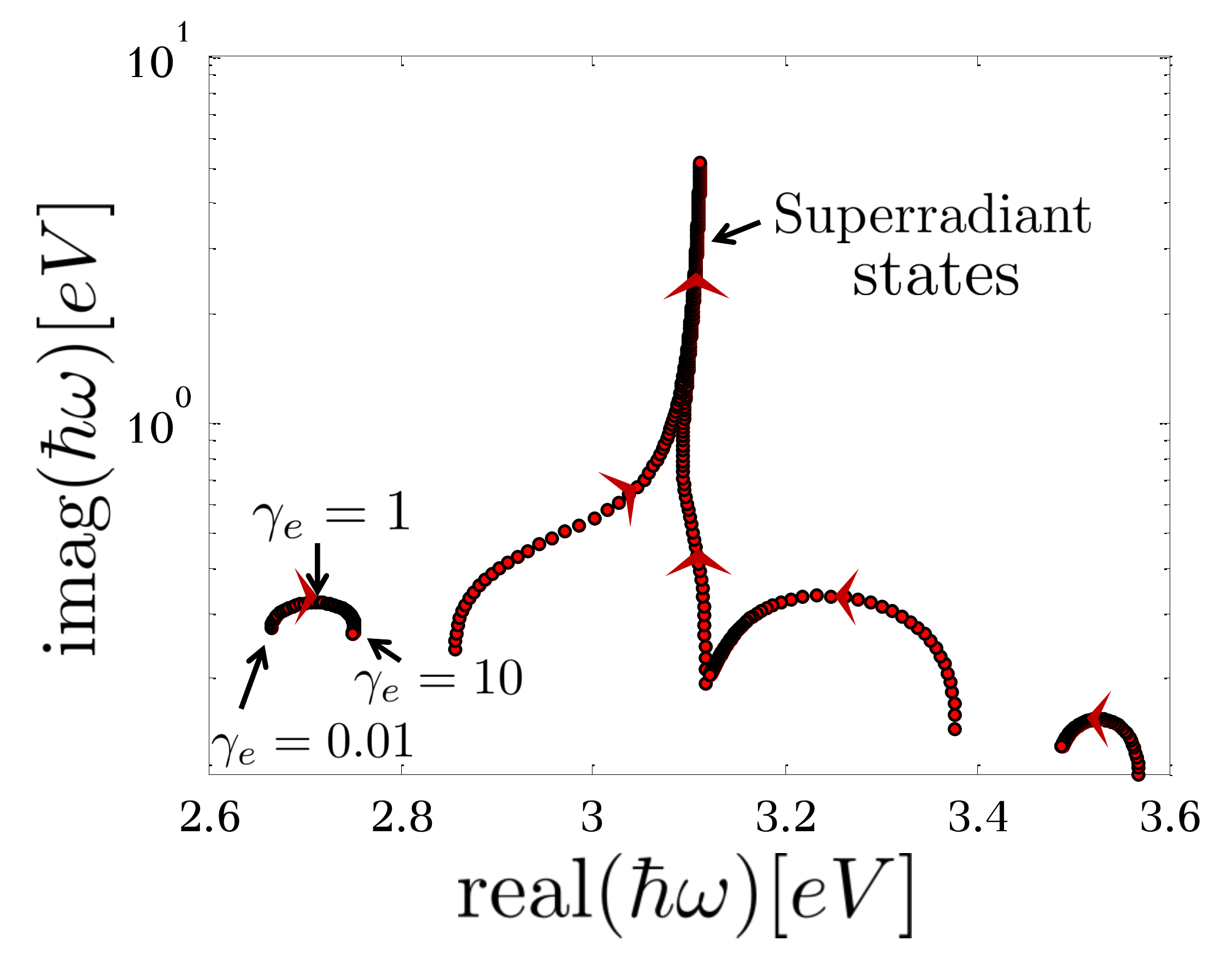}%
}

\captionsetup[subfigure]{justification=justified,singlelinecheck=false}
\caption{Complex-plane trajectories of the effective Hamiltonian for a one-dimensional chain of five identical silver spheres with radii of (a) 10 nm and (b) 40 nm. The spheres are in contact with one another and the dipoles are oriented along the waveguide. The arrows show the direction of the evolution as $\gamma_e$ changes from 0.01 eV to 10 eV.} \label{complex_eigenvals_sphrs}
\end{figure}

Through its coupling to the two probes, the system can undergo an additional superradiance phase transition, other than that discussed in the previous section. This is illustrated by considering two different plasmonic waveguides. In both cases, according to our findings in the last section, in order to maximize the coupling between the neighboring sites the spheres are in contact with one another and the dipole orientation of the spheres is considered to be along the waveguide (horizontal orientation). In the first case, a chain of five silver spheres with radii of 10 nm is considered (Fig. \ref{plasmonic_waveguide_schematics}). The resulting effective Hamiltonian (\ref{effective_hamiltonian_plasmonic_waveguide}) describing the system is a 5x5 square matrix with diagonal elements, $\omega_0=3.3468 + i0.0519$ eV, and off-diagonal matrix elements $\kappa=-0.2459 + i0.0029$ eV. The continuum coupling coefficient $\gamma_e$ is treated as a variable that changes from small, $\gamma_e=0.01$ eV, to extreme values $\gamma_e=10$ eV. The evolution of the complex eigenvalues of the effective Hamiltonian as the coupling to the continuum varies is shown in Fig. \ref{complex_eigenvals_sphrs}(a). At small values of $\gamma_e$ all the eigenvalues acquire a small width through the coupling to the continuum. The widths of the complex eigenmodes almost uniformly increase as the system is more strongly coupled to the continuum, up until $\gamma_e \approx 1$. At this point, the eigenvalues have reached their maximum width and, with further increasing $\gamma_e$, the system undergoes a phase transition (superradiance transition) when the eigenmodes become segregated into two distinct categories: superradiant and subradiant states. At strong coupling, the two superradiant states, their number being equal to the number of continuum channels (two probes), steal the entire available width of the system and leave the remaining states as narrow resonances.

The second waveguide differs only in that the size of the spheres now have radii of 40 nm. In this case, the diagonal unperturbed frequencies are $\omega_0=3.1172 + i0.1910$ eV and the off-diagonal coupling coefficients are $\kappa=-0.2606 + i0.0475$ eV. The continuum coupling $\gamma_e$ is again varied from $\gamma_e=0.01$ eV to $\gamma=10$ eV, and the complex eigenvalues are plotted in Fig. \ref{complex_eigenvals_sphrs}(b).  In general the picture is similar to the previous case. The superradiant transition can be clearly seen as the coupling $\gamma_e$ increases to extreme values.

We now study the propagation of a signal through the two waveguides by calculating the transmission coefficient. Using (\ref{ProcessAmplitude}) and (\ref{Transmission_amp}) we arrive at the following expression for the transmission coefficient
\begin{equation}
T(\hbar \omega_e) = \Bigg| \frac{\gamma_e/ \kappa}{ \prod_{r=1}^{N}\big[(\hbar \omega_e - \hbar \omega_r)/ \kappa \big]} \Bigg|^2,
\end{equation}
where $\omega_r$ are the complex frequencies of the effective Hamiltonian (\ref{effective_hamiltonian_plasmonic_waveguide}) and $N$ is its dimension.

\begin{figure}[h!]
\vspace{7mm}
\centering
\captionsetup{justification=centering}
\captionsetup[subfigure]{margin={1cm,0cm}}
\hspace{-2.25mm}
\subfloat[]{
  \includegraphics[height=4.3cm,width=6.15cm]{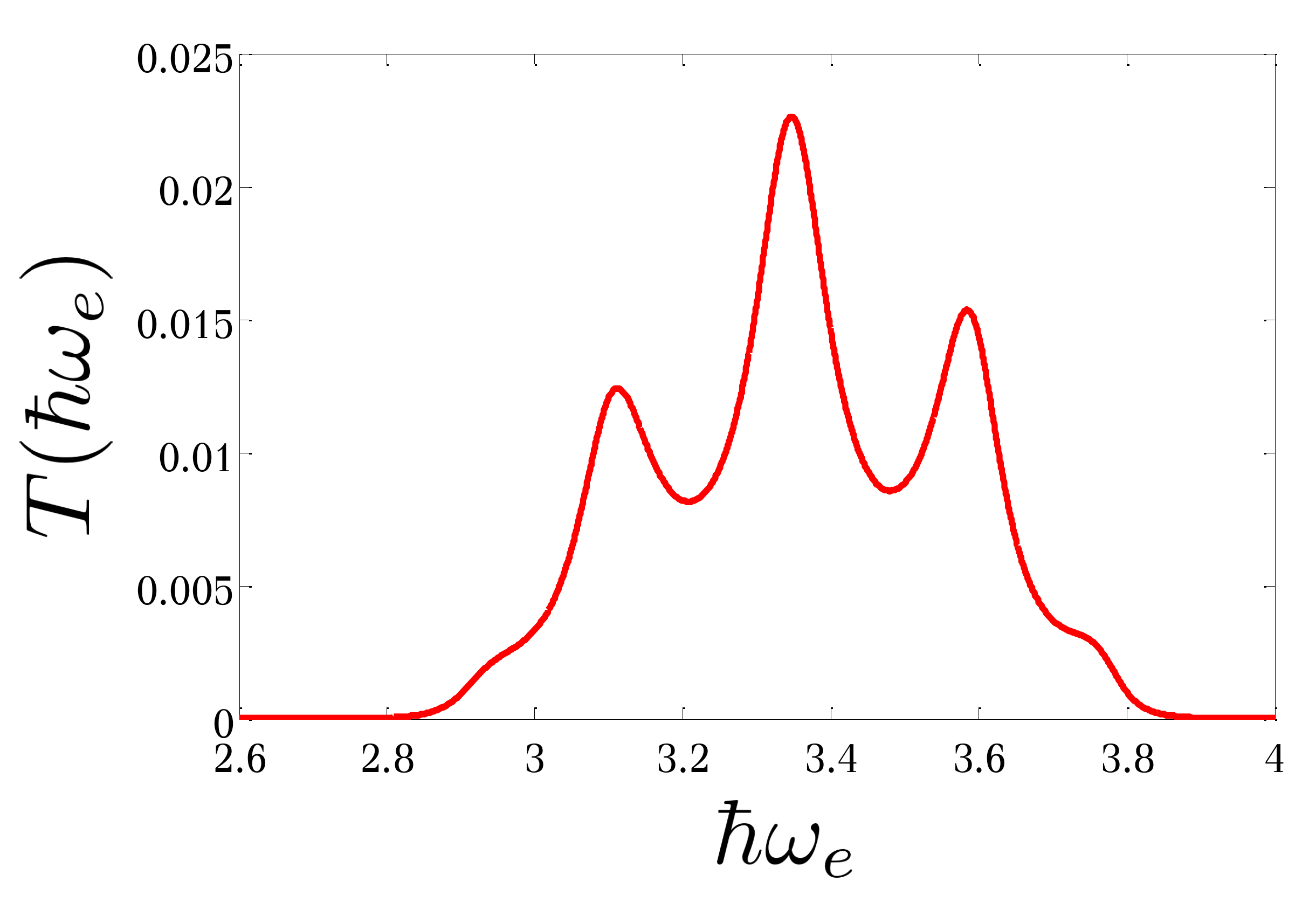}%
}

\captionsetup[subfigure]{margin={1cm,0cm}}
\hspace{0.5mm}
\subfloat[]{%
  \includegraphics[height=4.3cm,width=6cm]{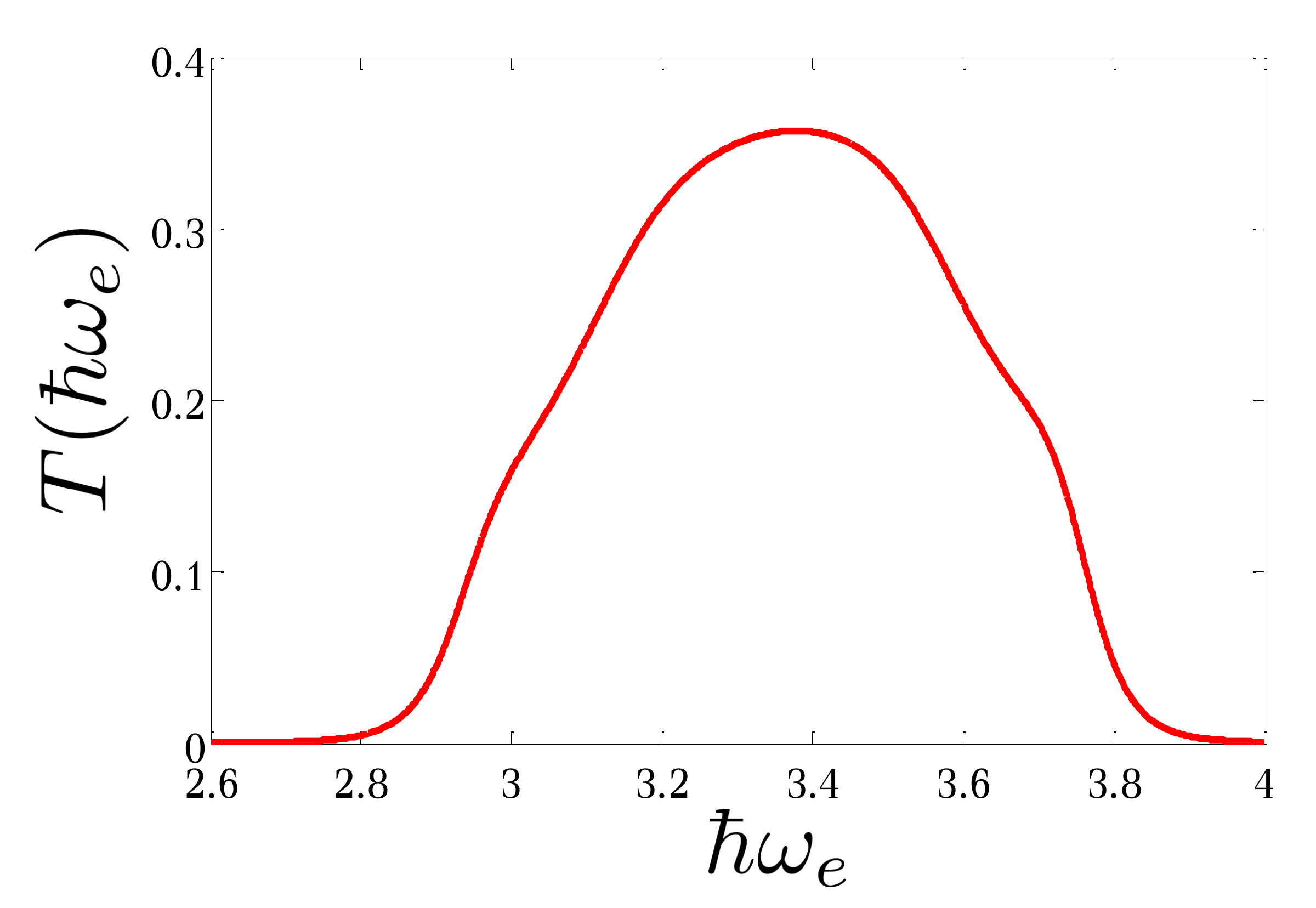}%
}

\captionsetup[subfigure]{margin={1cm,0cm}}
\hspace{-0.5mm}
\subfloat[]{%
  \includegraphics[height=4.3cm,width=6cm]{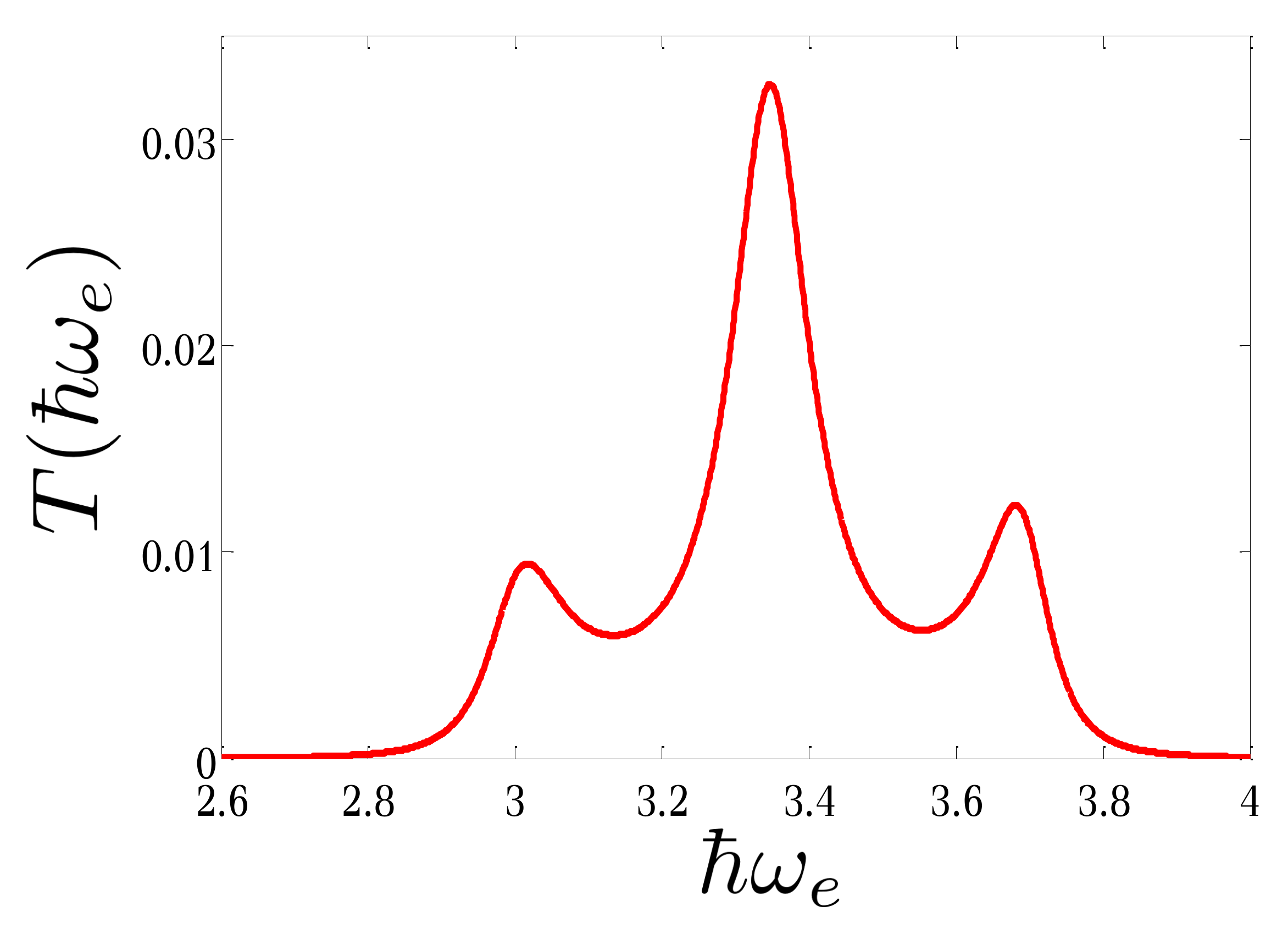}%
}
\captionsetup[subfigure]{justification=justified,singlelinecheck=false}
\caption{Transmission through a one-dimensional chain of five identical spheres with radii of 10 nm as a function of excitation source frequency $\omega_e$ for (a) $\gamma_e=$0.03 eV, (b) $\gamma_e=$0.55 eV, and (c) $\gamma_e=$10 eV.} \label{PlasmonicTransport10nm}
\end{figure}

Transmission as a function of the excitation frequency, $\omega_e$, is shown in Fig. \ref{PlasmonicTransport10nm} for the waveguide with 10 nm spheres. At weak coupling to the continuum, $\gamma_e=0.03$ eV, Fig. \ref{PlasmonicTransport10nm}(a), the five resonances are distinguishable. However, the resonances are not well separated due to the complex coupling coefficient between spheres, $\kappa$, which provide the eigenvalues of the effective Hamiltonian an initial width even for the closed system ($\gamma_e=0$). The case of intermediate coupling, when $\gamma_e=0.55$ eV, is shown in Fig. \ref{PlasmonicTransport10nm}(b). This is when the system is on the road to superradiance transition and all the eigenvalues of the Hamiltonian have large widths. Consequently, the resonances overlap and the transmission is dramatically enhanced. The case of strong couplings, Fig. \ref{PlasmonicTransport10nm}(c), has a picture similar to that of the weak coupling case. However, only three resonances remain. The two giant superradiant states do not participate in signal transmission and transmission is greatly suppressed due to the small width of the remaining subradiant states.

\begin{figure}[h!]
\centering
\captionsetup{justification=centering}
\captionsetup[subfigure]{margin={1cm,0cm}}
\hspace{-5.5mm}
\subfloat[]{
  \includegraphics[height=4.3cm,width=5.9cm]{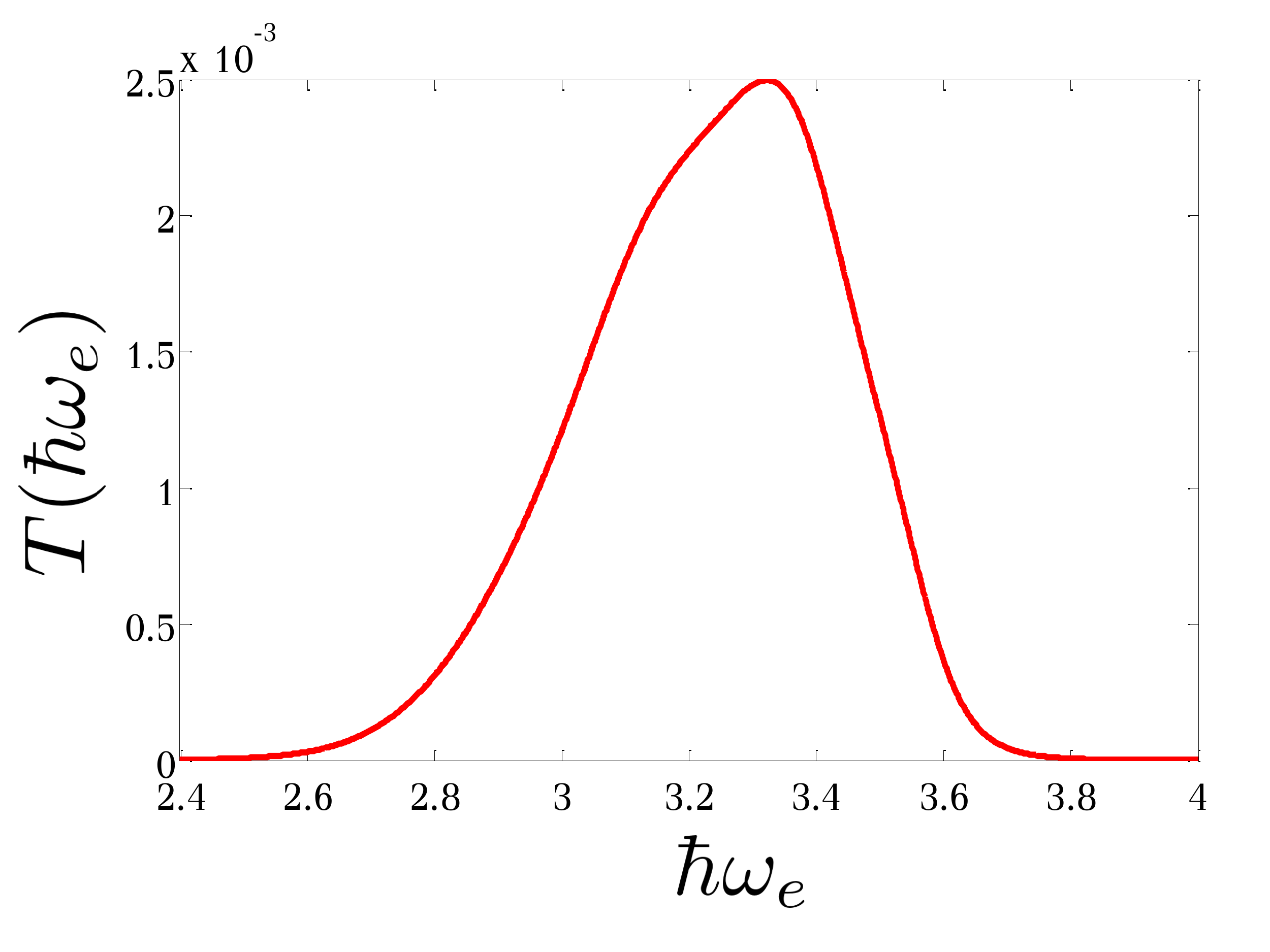}%
}

\captionsetup[subfigure]{margin={1cm,0cm}}
\hspace{-4.5mm}
\subfloat[]{%
  \includegraphics[height=4.3cm,width=6cm]{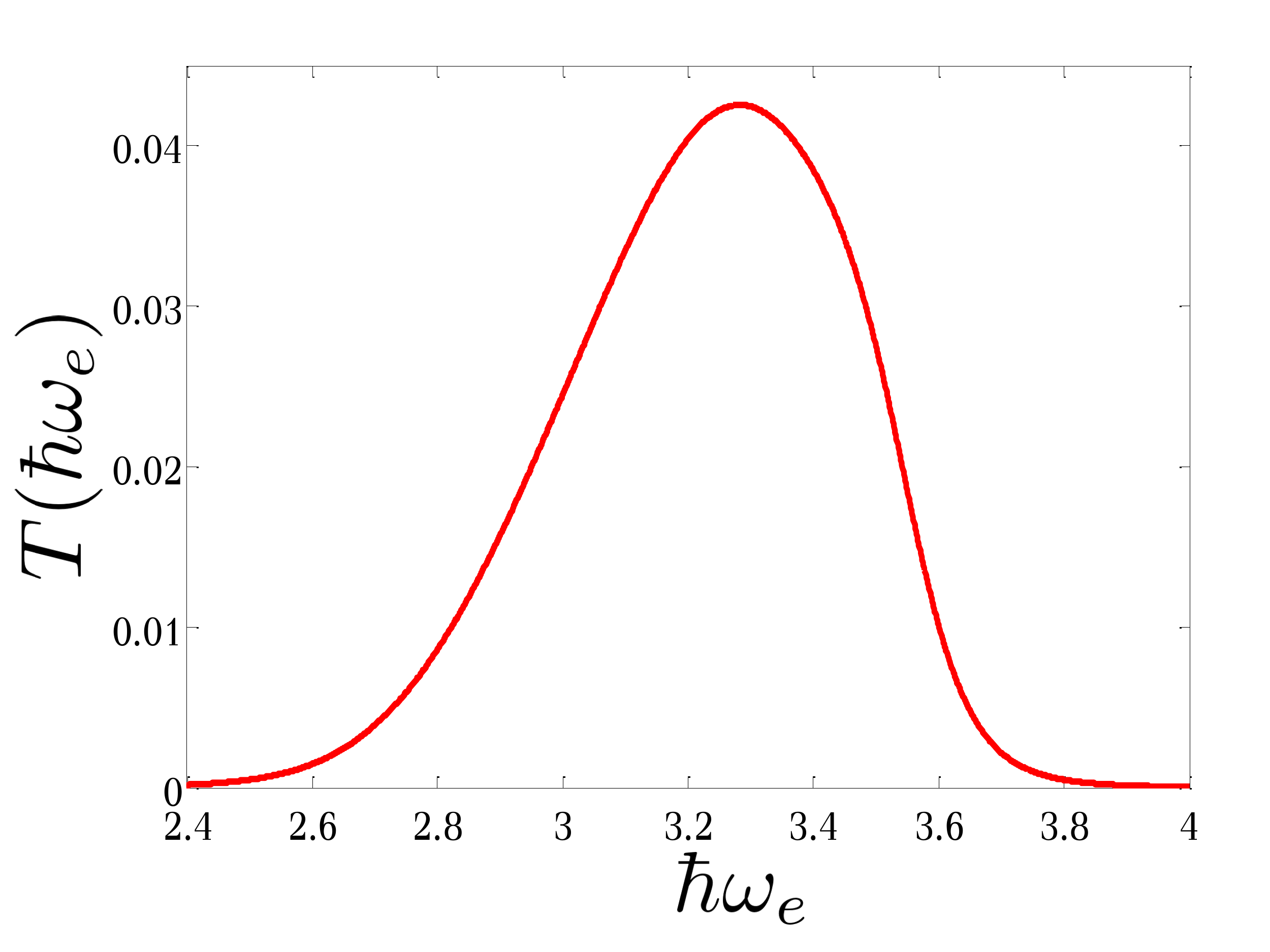}%
}

\captionsetup[subfigure]{margin={1cm,0cm}}
\hspace{-2.6mm}
\subfloat[]{%
  \includegraphics[height=4.3cm,width=5.84cm]{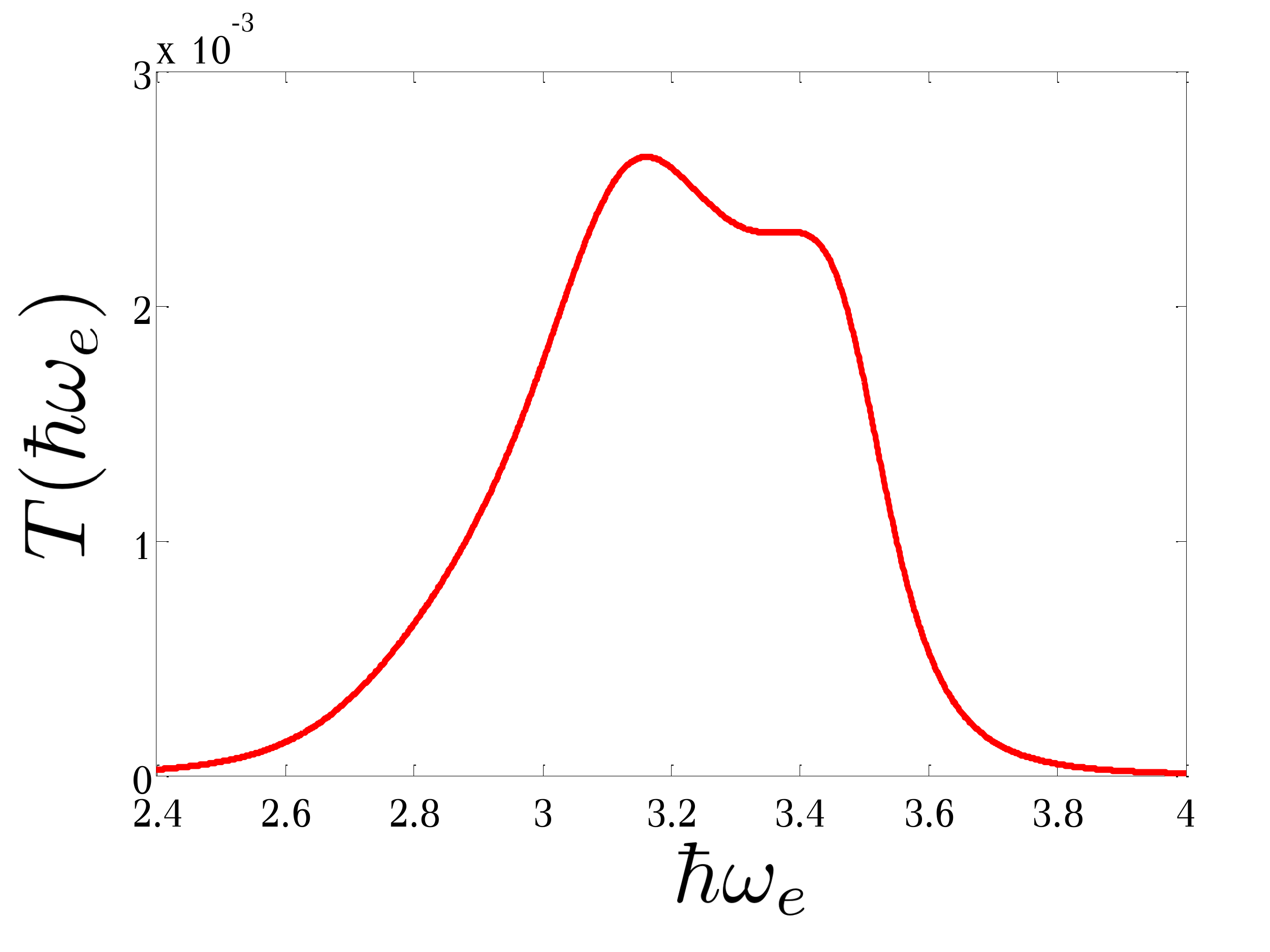}%
}
\captionsetup[subfigure]{justification=justified,singlelinecheck=false}
\caption{Transmission through a one-dimensional chain of five identical spheres with radii of 40 nm as a function of excitation source frequency $\omega_e$ for (a) $\gamma_e=$0.05 eV, (b) $\gamma_e=$0.7 eV, and (c) $\gamma_e=$10 eV.} \label{PlasmonicTransport40nm}
\end{figure}

We follow the same steps of weak, intermediate and strong coupling to continuum in order to study transmission through the waveguide with 40 nm spheres. Due to larger coupling, $\kappa$, between adjacent spheres the eigenvalues of the effective Hamiltonian poses a relatively large initial width even for small coupling to the continuum. Therefore, contrary to the previous case, the resonances overlap and are not separated even at weak coupling, Fig. \ref{PlasmonicTransport40nm}(a). Similar to before, the transmission is greatly enhanced at the superradiance transition, and at extreme couplings, we are back to suppressed transmission.

\section{Conclusion} \label{secVI}

We studied the resonant frequencies of plasmonic spherical nanoantennas by solving the full wave equation. These eigenfrequencies are always complex due to radiation and damping. Utilizing the effective non-Hermitian Hamiltonian framework, it was shown that a system of coupled two spheres can have modes with distinct properties; a superradiant mode with enhanced radiation and a dark mode with extremely damped radiation. 

Signal transmission through one dimensional chains was also considered. The coupling of the edge spheres to the continuum can drastically change transport properties of the system. A different superradiant transition arises through this interaction. Transmission is greatly enhanced at this transition. 

A possible direction to improve the accuracy of the results is to modify the Drude-Sommerfeld model by including terms that take into account surface scattering effects. It would be interesting to study the contribution of surface scattering to the total damping and radiation of the nanospheres. Another possibility is to consider higher order modes and their effect on the eigenfrequencies of the coupled system. These are left for future work.

\vspace{7mm}
\begin{acknowledgments}
A.~T. is grateful to his PhD advisor, Vladimir Zelevinsky for his guidance and many helpful discussions and thanks A.~Stain for her support and assistance. 

S. R. thanks his doctoral advisor J. Verboncoeur for his mentorship. 
\end{acknowledgments}

\appendix
\section{Quasi Normal Modes Normalization} \label{QNMNAp}
\begin{widetext}
In this appendix we explicitly normalize the fields of a plasmonic sphere by evaluating the normalization expression (\ref{normalization_int}) 
\begin{equation} \label{normalization_int2}
\int_V \sigma(\vec{r},\omega) \vec{\mathcal{E}}(\vec{r}).\vec{\mathcal{E}}(\vec{r}) d^3r + \frac{i\epsilon_{\text{out}}}{2k} \int_{\partial V} \vec{\mathcal{E}}(\vec{r}).\vec{\mathcal{E}}(\vec{r}) d^2r=1.
\end{equation}
Due to the homogeneity of the sphere dielectric function $\epsilon_{\text{in}}$, and the surrounding background $\epsilon_{\text{out}}$, the modified dielectric function $\sigma(\omega)$ given in (\ref{modified_dielectric}) is only a function of frequency and can be taken out of the integral. In what follows we first evaluate the volume term in (\ref{normalization_int2}) assuming the normalization volume itself is a sphere with a radius $R$ where $a \ll R$. We first consider the volume term. Using the field expressions given in (\ref{field_exprs}) the volume term of the normalization (\ref{normalization_int}) is expressed as
\begin{align} \label{volumeterm1}
& \sigma(\omega) \int_V  \ \vec{\mathcal{E}}(\vec{r}).\vec{\mathcal{E}}(\vec{r}) d^3r =\sigma(\omega) \ \zeta^2 \ \int_0^{R} dr r^2 \ C^2(r;a) \Bigg\{  \big(\ell(\ell+1)\big)^2 \bigg( \frac{f_{\ell}(kr)}{kr} \bigg)^2 \int d\Omega \bigg(Y^m_\ell(\theta,\phi)\bigg)^2  \nonumber \\
&+  \bigg( \frac{1}{kr} \frac{\partial}{\partial(kr)} \big( krf_{\ell}(kr) \big) \bigg)^2  \int d\Omega \bigg[  \bigg( \frac{\partial}{\partial \theta} Y_\ell^m(\theta,\phi) \bigg)^2 +\frac{1}{\text{sin}^2\theta} \bigg( \frac{\partial}{\partial \phi} Y_\ell^m(\theta,\phi) \bigg)^2 \bigg] \Bigg\},
\end{align}
where $d\Omega=\sin \theta d\theta d \phi$ is the solid angle differential in spherical coordinates. The integrals involving spherical harmonics can be evaluated by using the orthogonality relation of the tesseral harmonics and identity (\ref{SphericalHarmonic_angular_integral_identity}) in Appendix \ref{THandIs}. Thus, (\ref{volumeterm1}) reduces to
\begin{align} \label{volumeterm2}
& \sigma(\omega) \int_V  \ \vec{\mathcal{E}}(\vec{r}).\vec{\mathcal{E}}(\vec{r}) d^3r= \nonumber \\
& \sigma(\omega) \ \zeta^2 \ \ell(\ell+1) \int_0^R dr r^2 \ C^2(r;a) \Bigg\{  \ell(\ell+1) \bigg( \frac{f_{\ell}(kr)}{kr} \bigg)^2  
+  \bigg( \frac{1}{kr} \frac{\partial}{\partial(kr)} \big( krf_{\ell}(kr) \big) \bigg)^2   \Bigg\}.
\end{align}
Due to the discontinuity of the function $f_{\ell}(kr)$ and the coefficients $C(r;a)$ at the surface of the plasmonic sphere [see eqn. (\ref{general_constant})], the radial integral in (\ref{volumeterm2}) has to be divided into two terms: $\int_0^R=\int_0^a + \int_a^R$. Each term can be evaluated with the help of (\ref{SphericalHarmonic_integral_identity2}) and (\ref{SphericalHarmonic_integral_identity3}). This brings us to the final expression for the volume term of the normalization
\begin{equation} \label{volumeterm3}
\sigma(\omega) \int_V  \ \vec{\mathcal{E}}(\vec{r}).\vec{\mathcal{E}}(\vec{r}) d^3r = I \big[ j_{\ell}(k_{\text{in}}a) \big] -  I\big[h_{\ell}^{(2)}(k_{\text{out}}a)\big] + I\big[h_{\ell}^{(2)}(k_{\text{out}}R)\big],
\end{equation}
where the functional $I\big[f_{\ell}(kr)\big]$ is defined as
\begin{align} \label{volumeterm4}
& I\big[f_{\ell}(kr)\big]=  \sigma(\omega) \ \zeta^2 \ C^2(r;a)\frac{\ell(\ell+1)}{k^2} \bigg[ r f_{\ell}^2(kr)+kr^2 f_{\ell}(kr) f_{\ell}^{'}(kr)+\frac{k^2r^3}{2}\Big(f_{\ell}^2(kr)-f_{\ell-1}(kr)f_{\ell+1}(kr) \Big) \bigg].
\end{align}
As before, $f_{\ell}^{'}(kr)$ implies differentiation with respect to the argument i.e. $\frac{\partial}{\partial (kr)} f_{\ell}(kr)$. Note that in evaluating the first term of the right hand side of (\ref{volumeterm3}), $I \big[ j_{\ell}(k_{\text{in}}a) \big] $, one has to use all the parameters corresponding to the region interior to the plasmonic sphere. i.e. $\epsilon_{\text{in}}(\omega)$, $k_{\text{in}}$ and $C(r;a)$ for $r \leq a$ as it is defined in (\ref{general_constant}). Accordingly, the same applies to the second and third terms, $I\big[h_{\ell}^{(2)}(k_{\text{out}}a)\big]$ and $I\big[h_{\ell}^{(2)}(k_{\text{out}}R)\big]$, for which one has to use the parameters corresponding to the background material. 

Next we show that the last term in the right hand side of (\ref{volumeterm3}), $I\big[h_{\ell}^{(2)}(k_{\text{out}}R)\big]$, exactly cancels out with the surface term in (\ref{normalization_int}). Therefore, as expected, the normalization condition becomes independent of the integration volume. Because the integration sphere was assumed to be sufficiently large, asymptotic expressions can be used to evaluate both these terms. The spherical Hankel functions at large radial distances have the following asymptotic form
\begin{equation} \label{Asymptotic_hankel}
h_{\ell}^{(2)}(kr)\sim i^{(\ell+1)} \frac{e^{-ikr}}{kr} \Bigg(1-i \frac{\ell(\ell+1)}{2kr} \Bigg).
\end{equation}
The last term in the right hand side of (\ref{volumeterm4}) can therefore be approximated as
\begin{equation} \label{VolumeLargeContr}
I\big[h_{\ell}^{(2)}(k_{\text{out}}R)\big] \approx \frac{-i \epsilon_{\text{out}}}{2k} C^2(r;a) (-)^{\ell+1}e^{-2ik_{\text{out}}R}.
\end{equation}
We now consider the surface term of the normalization condition (\ref{normalization_int}). Considering that in the far field the dominant terms are $\mathcal{E}_\theta$ and $\mathcal{E}_\phi$ and using the asymptotic form of the spherical Hankel function (\ref{Asymptotic_hankel}), the leading order of the surface term is
\begin{equation} \label{Surfaceterm_final}
 \frac{i \epsilon_{\text{out}}}{2k} \int_{\partial V} \vec{\mathcal{E}}(\vec{r}).\vec{\mathcal{E}}(\vec{r}) d^2r \approx  \frac{i \epsilon_{\text{out}}}{2k^3} C^2(r;a) (-)^{\ell+1} e^{-2ik_{\text{out}}R}
\end{equation}

It is clear that the surface term (\ref{Surfaceterm_final}) and (\ref{VolumeLargeContr}) of the volume term cancel out. With this, the normalization condition (\ref{normalization_int2}) reduces to
\begin{equation} \label{normalization_reduced}
I \big[ j_{\ell}(k_{\text{in}}a) \big] -  I\big[h_{\ell}^{(2)}(k_{\text{out}}a)\big]=1,
\end{equation}
which provide us the coefficient $\zeta$ in (\ref{field_exprs}). 
\end{widetext}

\section{Tesseral Harmonics and Spherical Functions Identities} \label{THandIs}
This appendix contains the definition of tesseral harmonics and a list of identities that are used throughout the paper.

The tesseral harmonic are linear superpositions of the complex spherical harmonics with same $\ell$ and opposite sign $m$ values. Therefore the azimuthal dependency of the functions are in the form of $\text{sin}(m\phi)$ and $\text{cos}(m\phi)$ instead of the usual exponential form $e^{im\phi}$. They are defined as \cite{THarmonics}:

\begin{align}
Y_{\ell}^m(\theta,\phi) = \begin{cases}
\sqrt{\frac{2\ell+1}{2 \pi}\frac{(\ell-|m|)!}{(\ell+|m|)!}} P_{\ell}^{|m|}(\text{cos} \theta) \text{sin}(|m|\phi) & m<0 \\
\sqrt{\frac{2\ell+1}{4 \pi}} P_{\ell}^{m}  & m=0\\
\sqrt{\frac{2\ell+1}{2 \pi}\frac{(\ell-m)!}{(\ell+m)!}} P_{\ell}^{m}(\text{cos} \theta) \text{cos}(m\phi) & m>0
\end{cases}
\end{align}
where $P_{\ell}^{m}$ are the associated Legendre polynomials. The tesseral harmonics satisfy the same orthogonality relation as the complex spherical harmonics.

The tesseral harmonics also satisfy the following identity 

\begin{align} \label{SphericalHarmonic_angular_integral_identity}
& \int d\Omega \Bigg\{  \bigg( \frac{\partial}{\partial \theta} Y_\ell^m(\theta,\phi)  \bigg)^2 +\frac{1}{\sin^2\theta} \bigg( \frac{\partial}{\partial \phi} Y_\ell^m(\theta,\phi) \bigg)^2 \Bigg\} = \nonumber \\
& \ell (\ell+1).
\end{align}

This can be proven by starting with the fact that the harmonics fulfill the identity $r^2 \nabla^2 Y_{\ell}^m(\theta,\phi)=-\ell(\ell+1)Y_{\ell}^m(\theta,\phi)$. One can then get to (\ref{SphericalHarmonic_angular_integral_identity}), by calculating the matrix element of the operator $r^2 \nabla^2$ and using integration by parts. 

A useful identity of the spherical Bessel and Hankel functions is the following 

\begin{align} \label{SphericalHarmonic_integral_identity2}
\int dr \Bigg\{ \ell (\ell+1) f_{\ell}^2(kr) + \bigg( \frac{\partial}{\partial (kr)} krf_{\ell}(kr) \bigg)^2 \Bigg\}= \nonumber \\
r f_{\ell}^2(kr)+kr^2 f_{\ell}(kr) f_{\ell}^{'}(kr)+k^2 \int dr r^2 f_{\ell}^2(kr).
\end{align}
where as previously $f_{\ell}^{'}(kr)$ implies differentiation with respect to the argument. The equality can be proven by using  integration by parts and the spherical Bessel differential equation to get
\begin{align} 
& \ell (\ell+1) f_{\ell}^2(kr) = \nonumber  \\
 & k^2r^2 f_{\ell}(kr) f_{\ell}^{''}(kr)+2krf_{\ell}(kr)f_{\ell}^{'}(kr)+k^2r^2f_{\ell}^2(kr),
\end{align}
 and 
\begin{equation} \label{SphericalHarmonic_integral_identity3}
\int dr r^2  f_{\ell}^2(kr) = \frac{r^3}{2} \bigg(  f_{\ell}^2(kr)- f_{\ell-1}(kr) f_{\ell+1}(kr)  \bigg).
\end{equation}
\bibliography{Refs}

\end{document}